\documentclass[aps,prb,numerical,reprint,showkeys,longbibliography]{revtex4-2}
\usepackage{epsfig}
\usepackage{amsmath}
\usepackage{dcolumn}
\usepackage{color}
\begin{document}

% input definitions
%\input{rmmcoms}
%
%Commands for lecture notes and book

\newcommand{\VI}{Volume I}
\newcommand{\VII}{Volume II}

\newcommand{\eg}{{\it e.g.,\ }}
\newcommand{\ie}{{\it i.e.,\ }}
\newcommand{\etal}{{\it et al.}}
\newcommand{\etc}{{\it etc.\ }}
\newcommand{\via}{{\it via}}
\newcommand{\ai}{{\it ab initio}}
\newcommand{\Ai}{{\it Ab initio}}
\newcommand{\AI}{{\it Ab Initio}}

\newcommand{\wrt}{with respect to}
\newcommand{\rhs}{right hand side}

\newcommand{\kbt}{{\rm k_B T}}

\newcommand{\mc}{Monte Carlo}
\newcommand{\qmc}{quantum Monte Carlo}
\newcommand{\Qmc}{Quantum Monte Carlo}
\newcommand{\fn}{fixed-node}

\newcommand{\ham}{hamiltonian}
\newcommand{\wf}{wave function}
\newcommand{\dm}{density matrix}

\newcommand{\mb}{many-body}
\newcommand{\ip}{independent-particle}
\newcommand{\ipa}{independent-particle approximation}

\newcommand{\dft}{density functional theory}
\newcommand{\Dft}{Density functional theory}
\newcommand{\DFT}{Density Functional Theory}
\newcommand{\df}{density functional}
\newcommand{\Df}{Density functional}
\newcommand{\ld}{local density}
\newcommand{\lda}{local density approximation}
\newcommand{\ggr}{generalized gradient}
\newcommand{\gga}{generalized gradient approximation}
\newcommand{\KS}{Kohn-Sham}
\newcommand{\HK}{Hohenberg-Kohn}
\newcommand{\BO}{Born-Oppenheimer}
\newcommand{\CP}{Car-Parrinello}
\newcommand{\HF}{Hartree-Fock}
\newcommand{\HeF}{Hellmann-Feynman}
\newcommand{\ft}{force theorem}
\newcommand{\st}{stress theorem}
\newcommand{\Schr}{Schr\"{o}dinger}
\newcommand{\schro}{Schr\"{o}dinger}
\newcommand{\AM}{Ashcroft and Mermin}

\newcommand{\ex}{exchange}
\newcommand{\Ex}{Exchange}
\newcommand{\exc}{exchange-correlation}
\newcommand{\Exc}{Exchange-correlation}
\newcommand{\corr}{correlation}
\newcommand{\Corr}{Correlation}

% "\nrho" is to be able to change from n to rho
% for the density
\newcommand{\nrho}{n}

\newcommand{\Tip}{T_{s}}

\newcommand{\fcc}{fcc}
\newcommand{\bcc}{bcc}

\newcommand{\bad}{bond angle distribution function}
\newcommand{\bads}{bond angle distribution functions}
\newcommand{\eos}{equation of state}
\newcommand{\Nose}{Nos\'{e}}
\newcommand{\psp}{pseudopotential}
\newcommand{\psps}{pseudopotentials}
\newcommand{\Psp}{Pseudopotential}
\newcommand{\Psps}{Pseudopotentials}
\newcommand{\pswf}{pseudo-wave function}
\newcommand{\rdf}{radial distribution function}
\newcommand{\rdfs}{radial distribution functions}

% physics symbols

\newcommand{\bra}[1]         {\ensuremath{\langle #1|}}
\newcommand{\ket}[1]         {\ensuremath{|#1\rangle}}
\newcommand{\ev}[1]          {\ensuremath{\langle #1 \rangle}}

\newcommand{\adag}           {\ensuremath{a^{\dag}}}
\newcommand{\cdag}           {\ensuremath{c^{\dag}}}
% math symbols

%\newcommand{\bDelta}{\mbox{\boldmath $\Delta$}}
\newcommand{\bDelta}{\mbox{$\bf \Delta$}}
\newcommand{\bG}{{\bf G}}
\newcommand{\bGp}{\mbox{$\bG^{\prime}$}}
\newcommand{\bk}{{\bf k}}
\newcommand{\bnabla}{\mbox{\boldmath $\nabla$}}
\newcommand{\bq}{{\bf q}}
\newcommand{\bR}{{\bf R}}
\newcommand{\br}{{\bf r}}
\newcommand{\brp}{\mbox{$\br^{\prime}$}}
\newcommand{\bx}{{\bf x}}
\newcommand{\half}{{1 \over 2}}

\newcommand{\rij}{\mbox{${r_{ij}}$}}
\newcommand{\modrij}{\mbox{${|{\bf r}_{i} - {\bf r}_{j}|}$}}
\newcommand{\modrijt}{\mbox{${|\tilde{{\bf r}}_{i} - \tilde{{\bf r}}_{j}|}$}}
\newcommand{\modrrp}{\mbox{${|{\bf r} - {\bf r}^{\prime}|}$}}
\newcommand{\modriI}{\mbox{${|{\bf r}_{i} - {\bf R}_{I}|}$}}
\newcommand{\modRIJ}{\mbox{${|{\bf R}_{I} - {\bf R}_{J}|}$}}

\newcommand{\rmb}{\mbox{$\{ \vecr_{i} \} $}}
\newcommand{\Rmb}{\mbox{$\{ \vecR_{I} \} $}}

\newcommand{\rp}{\mbox{${r^{\prime}}$}}
\newcommand{\vecrp}{\mbox{${{\bf r}^{\prime}}$}}
\newcommand{\sigmap}{\mbox{${\sigma^{\prime}}$}}
\newcommand{\vecri}{\mbox{${{\bf r}_i}$}}
\newcommand{\vecRi}{\mbox{${{\bf R}_i}$}}
\newcommand{\vecRI}{\mbox{${{\bf R}_I}$}}
\newcommand{\vecrj}{\mbox{${{\bf r}_j}$}}
\newcommand{\vecRj}{\mbox{${{\bf R}_j}$}}
\newcommand{\vecRJ}{\mbox{${{\bf R}_J}$}}

% procedural macros
\newtheorem{prop}{Property}

\newcommand{\refapp}[1]{App.~\ref{#1}}
\newcommand{\refApp}[1]{Appendix~\ref{#1}}
\newcommand{\refchap}[1]{Chap.~\ref{#1}}
\newcommand{\refChap}[1]{Chapter~\ref{#1}}
\newcommand{\refeq}[1]{Eq.~\ref{#1}}
\newcommand{\refEq}[1]{Equation~\ref{#1}}
\newcommand{\reffig}[1]{Fig.~\ref{#1}}
\newcommand{\refFig}[1]{Figure~\ref{#1}}
\newcommand{\refsec}[1]{Sec.~\ref{#1}}
\newcommand{\refSec}[1]{Section~\ref{#1}}
\newcommand{\reftab}[1]{table~\ref{#1}}
\newcommand{\refTab}[1]{Table~\ref{#1}}

\newcommand{\refapps}[1]{Apps.~\ref{#1}}
\newcommand{\refApps}[1]{Appendices~\ref{#1}}
\newcommand{\refchaps}[1]{Chaps.~\ref{#1}}
\newcommand{\refChaps}[1]{Chapters~\ref{#1}}
\newcommand{\refeqs}[1]{Eqs.~\ref{#1}}
\newcommand{\refEqs}[1]{Equations~\ref{#1}}
\newcommand{\reffigs}[1]{Figs.~\ref{#1}}
\newcommand{\refFigs}[1]{Figures~\ref{#1}}
\newcommand{\refsecs}[1]{Secs.~\ref{#1}}
\newcommand{\refSecs}[1]{Sections~\ref{#1}}
\newcommand{\reftabs}[1]{tables~\ref{#1}}
\newcommand{\refTabs}[1]{Tables~\ref{#1}}

% letter abbreviations

\newcommand{\cA}{{\cal A}}
\newcommand{\cB}{{\cal B}}
\newcommand{\cC}{{\cal C}}
\newcommand{\cD}{{\cal D}}
\newcommand{\cE}{{\cal E}}
\newcommand{\cF}{{\cal F}}
\newcommand{\cG}{{\cal G}}
\newcommand{\cH}{{\cal H}}
\newcommand{\cI}{{\cal I}}
\newcommand{\cJ}{{\cal J}}
\newcommand{\cK}{{\cal K}}
\newcommand{\cL}{{\cal L}}
\newcommand{\cM}{{\cal M}}
\newcommand{\cN}{{\cal N}}
\newcommand{\cO}{{\cal O}}
\newcommand{\cP}{{\cal P}}
\newcommand{\cQ}{{\cal Q}}
\newcommand{\cR}{{\cal R}}
\newcommand{\cS}{{\cal S}}
\newcommand{\cT}{{\cal T}}
\newcommand{\cU}{{\cal U}}
\newcommand{\cV}{{\cal V}}
\newcommand{\cW}{{\cal W}}
\newcommand{\cX}{{\cal X}}
\newcommand{\cY}{{\cal Y}}
\newcommand{\cZ}{{\cal Z}}

\newcommand{\Hhat}{{\hat{H}}}
\newcommand{\rhohat}{{\hat{\rho}}}

\newcommand{\bfa}{{\bf a}}
\newcommand{\bfb}{{\bf b}}
\newcommand{\bfc}{{\bf c}}
\newcommand{\bfd}{{\bf d}}
\newcommand{\bfe}{{\bf e}}
\newcommand{\bff}{{\bf f}}
\newcommand{\bfg}{{\bf g}}
\newcommand{\bfh}{{\bf h}}
\newcommand{\bfi}{{\bf i}}
\newcommand{\bfj}{{\bf j}}
\newcommand{\bfk}{{\bf k}}
\newcommand{\bfl}{{\bf l}}
\newcommand{\bfm}{{\bf m}}
\newcommand{\bfn}{{\bf n}}
\newcommand{\bfo}{{\bf o}}
\newcommand{\bfp}{{\bf p}}
\newcommand{\bfq}{{\bf q}}
\newcommand{\bfr}{{\bf r}}
\newcommand{\bfs}{{\bf s}}
\newcommand{\bft}{{\bf t}}
\newcommand{\bfu}{{\bf u}}
\newcommand{\bfv}{{\bf v}}
\newcommand{\bfw}{{\bf w}}
\newcommand{\bfx}{{\bf x}}
\newcommand{\bfy}{{\bf y}}
\newcommand{\bfz}{{\bf z}}
\newcommand{\bfA}{{\bf A}}
\newcommand{\bfB}{{\bf B}}
\newcommand{\bfC}{{\bf C}}
\newcommand{\bfD}{{\bf D}}
\newcommand{\bfE}{{\bf E}}
\newcommand{\bfF}{{\bf F}}
\newcommand{\bfG}{{\bf G}}
\newcommand{\bfH}{{\bf H}}
\newcommand{\bfI}{{\bf I}}
\newcommand{\bfJ}{{\bf J}}
\newcommand{\bfK}{{\bf K}}
\newcommand{\bfL}{{\bf L}}
\newcommand{\bfM}{{\bf M}}
\newcommand{\bfN}{{\bf N}}
\newcommand{\bfO}{{\bf O}}
\newcommand{\bfP}{{\bf P}}
\newcommand{\bfQ}{{\bf Q}}
\newcommand{\bfR}{{\bf R}}
\newcommand{\bfS}{{\bf S}}
\newcommand{\bfT}{{\bf T}}
\newcommand{\bfU}{{\bf U}}
\newcommand{\bfV}{{\bf V}}
\newcommand{\bfW}{{\bf W}}
\newcommand{\bfX}{{\bf X}}
\newcommand{\bfY}{{\bf Y}}
\newcommand{\bfZ}{{\bf Z}}

\newcommand{\veca}{{{\bf a}}}
\newcommand{\vecb}{{{\bf b}}}
\newcommand{\vecc}{{{\bf c}}}
\newcommand{\vecd}{{{\bf d}}}
\newcommand{\vece}{{{\bf e}}}
\newcommand{\vecf}{{{\bf f}}}
\newcommand{\vecg}{{{\bf g}}}
\newcommand{\vech}{{{\bf h}}}
\newcommand{\veci}{{{\bf i}}}
\newcommand{\vecj}{{{\bf j}}}
\newcommand{\veck}{{{\bf k}}}
\newcommand{\vecl}{{{\bf l}}}
\newcommand{\vecm}{{{\bf m}}}
\newcommand{\vecn}{{{\bf n}}}
\newcommand{\veco}{{{\bf o}}}
\newcommand{\vecp}{{{\bf p}}}
\newcommand{\vecq}{{{\bf q}}}
\newcommand{\vecr}{{{\bf r}}}
\newcommand{\vecs}{{{\bf s}}}
\newcommand{\vect}{{{\bf t}}}
\newcommand{\vecu}{{{\bf u}}}
\newcommand{\vecv}{{{\bf v}}}
\newcommand{\vecw}{{{\bf w}}}
\newcommand{\vecx}{{{\bf x}}}
\newcommand{\vecy}{{{\bf y}}}
\newcommand{\vecz}{{{\bf z}}}
\newcommand{\vecA}{{{\bf A}}}
\newcommand{\vecB}{{{\bf B}}}
\newcommand{\vecC}{{{\bf C}}}
\newcommand{\vecD}{{{\bf D}}}
\newcommand{\vecE}{{{\bf E}}}
\newcommand{\vecF}{{{\bf F}}}
\newcommand{\vecG}{{{\bf G}}}
\newcommand{\vecH}{{{\bf H}}}
\newcommand{\vecI}{{{\bf I}}}
\newcommand{\vecJ}{{{\bf J}}}
\newcommand{\vecK}{{{\bf K}}}
\newcommand{\vecL}{{{\bf L}}}
\newcommand{\vecM}{{{\bf M}}}
\newcommand{\vecN}{{{\bf N}}}
\newcommand{\vecO}{{{\bf O}}}
\newcommand{\vecP}{{{\bf P}}}
\newcommand{\vecQ}{{{\bf Q}}}
\newcommand{\vecR}{{{\bf R}}}
\newcommand{\vecS}{{{\bf S}}}
\newcommand{\vecT}{{{\bf T}}}
\newcommand{\vecU}{{{\bf U}}}
\newcommand{\vecV}{{{\bf V}}}
\newcommand{\vecW}{{{\bf W}}}
\newcommand{\vecX}{{{\bf X}}}
\newcommand{\vecY}{{{\bf Y}}}
\newcommand{\vecZ}{{{\bf Z}}}

\newcommand{\vectau}{{{\bf \tau}}}
\newcommand{\veckappa}{{{\bf \kappa}}}

\newcommand{\bftau}{\bf {\tau}}
% **************************************************

\title{Energy density and stress fields in quantum systems }
\author{Richard M.~Martin}
\affiliation{Department of Physics, University of Illinois at
         Urbana-Champaign, Urbana, Illinois 61801, USA}
\affiliation{Department of Applied Physics, Stanford University,
        Stanford, California 94305, USA}
\author{Nithaya Chetty}
\affiliation{School of Physics, University of the Witwatersrand, Johannesburg, South Africa}
\author{Dallas R. Trinkle}
\affiliation{Department of Materials Science and Engineering, University of Illinois at
         Urbana-Champaign, Urbana, Illinois 61801, USA}
\date{\today}
%\maketitle

\begin{abstract}
There has been an enduring interest and controversy about whether or not one can define physically meaningful energy density and stress fields, $e(\textbf{r})$ and $\sigma_{\alpha \beta}(\textbf{r})$ in quantum systems.
A key issue is the kinetic energy since the well-known forms,
$\frac{1}{2}|\nabla \Psi|^2$ and  $-\frac{1}{2}\Psi\nabla^2 \Psi$, lead to different densities, and analogous issues arise for interaction energy terms. This paper considers the ground state of a system of many interacting particles in an external potential, and presents a resolution to the problems in steps. 1) For the kinetic energy all effects of exchange and correlation are shown to be unique functions defined at each point $\textbf{r}$; all issues of nonuniqueness are relegated to terms that involve only the density $n(\textbf{r})$ and are equivalent to an effective single-particle problem with ground state wave function $s(\textbf{r}) = \sqrt{n(\textbf{r})/N}$. 
2) Interactions can be considered in two ways: in terms of potentials acting on particles or in terms of the interaction fields, \eg the Maxwell form in terms of electric fields. In each case, the problem reduces to a mean field part that is a function of the density and a part due to correlation that is uniquely defined; however, it is different for the two cases. 3) The final results follow from the nature of energy and stress. 
Because the energy determines the ground state itself through the variational
principle, the energy density approach leads directly to kinetic energy in the form $-\frac{1}{2}s\nabla^2 s$ and interactions in terms of potentials acting on the particles.
This leads
naturally to density functional theory and provides an interpretation in which the energy density $e(\textbf{r})$ is equilibrated to minimize
fluctuations with the same chemical potential at all points $\textbf{r}$.  
On the other hand, stress is related to forces, and the only physically acceptable
expressions for the stress field involve the combination $\frac{1}{2}[s\nabla^2 s - |\nabla s|^2]$, as derived by \Schr, Pauli and others, 
and Coulomb interactions in terms of electric fields, not potentials. 
Together these results lead to well-defined formulations of energy density and stress
fields that are physically motivated and based on a clear set of arguments.
\end{abstract}

  %CHECK!

\maketitle

%\marginpar{Modified abstract & added last sentence May 21}
\section{Introduction} 
\label{sec:Introduction}

Since the early days of quantum mechanics there has been great interest and controversy
concerning the possibility - or impossibility - of finding unique,
well-defined, physically meaningful expressions for energy density\cite{schr27,fock30,pauli33,epstein-10-1063,che92,hammer95,rapcewicz98,MHCohen00,ernzerhof02,martin04-20,anderson-JCPA114-8884,yu2011} and stress fields\cite{schr27,fock30,pauli33,epstein-10-1063,irving50,lan59,kugler67,schofield82,RMM-nielsen85,godfrey88,wajnryb95,grafenstein96,filippetti00,rogers02,martin04-20,anderson-JCPA114-8884,guevara-in-book}
defined at each point $\vecr$ in quantum systems.
References to many classic papers
%starting in the early days of quantum mechanics
are given in Refs. \cite{anderson-JCPA114-8884} and \cite{RMM-nielsen85}.
There are good reasons to conclude that such local quantities
can never be unique: In classical physics 
the interaction energy can be assigned among the interacting
particles or to the space between the particles, without changing
any observables; Coulomb interactions can be treated as potentials acting on particles or electric fields between particles\cite{lan60,jac62}.  At first sight, it appears that quantum
mechanics adds further nonuniqueness.  Whereas classical kinetic
energies are directly associated with the particles and their
velocities at definite positions, in quantum mechanics there are
different choices for the kinetic energy operator, $\frac{1}{2}|\nabla \Psi|^2$,   $-\frac{1}{2}\Psi\nabla^2 \Psi$  or a linear combination,  which result in
different kinetic energy densities.

Nevertheless, there has been much work to use such concepts to interpret chemical bonding and other phenomena.
The paper by Anderson and coworkers\cite{anderson-JCPA114-8884} ``How Ambiguous Is the Local Kinetic Energy?'' has extensive references the literature and it emphasizes the role of the stress field as well as energy density to characterize chemical bonding.  Examples include the reviews 
``The Physical Nature of the Chemical Bond'' \cite{ruedenberg-RevModPhys.34.326} and ``Electronic Stress as a Guiding Force
for Chemical Bonding'' \cite{guevara-in-book}, the ``Atoms in Molecules'' analysis of Bader\cite{bader90,bader91}, calculations of energies for selected regions\cite{che92,yu2011}, interpretation of DFT calculations\cite{MHCohen00}, ``DFT-chemical pressure analysis'' \cite{fredrickson2012},  ``Quantum pressure focusing'' analysis of shell structure and bonding\cite{tao-PhysRevLett.100.206405,tsirelson:px5014}, and many other works, \eg\ as cited in \cite{anderson-JCPA114-8884}, \cite{fredrickson2012} and a more recent review \cite{kohyama-2021MT-M2020291}. A perceptive description of the interpretation of bonding in terms of stresses is in the undergraduate thesis of Feynman which is available online\cite{feynman39a}.  The use of densities to interpret properties of materials is not the purpose of the present work, although it is relevant to make contact with  the literature to bring out physical principles.  
% Commented out Dec. 5 \marginpar{Add examples in  Fredrickson\cite{fredrickson2012}, Parr, others?}

 The purpose of this paper is to provide formulations of energy density and stress fields that are physically motivated and based on a clear set of arguments. Energy density and stress (or local pressure) have the same units, energy per unit volume, and they both are expressed in terms of kinetic operators and interactions, but they are fundamentally different.  A thesis of the present work is that we must examine carefully the similarities and differences, and the way they are used in order to derive unique expressions and to clarify the reasons for different choices for the kinetic operators and the way interactions are accounted for. The analysis applies for the ground state of a general many-body systems described by the nonrelativistic \Schr\ equation, which is sufficient for most aspects of quantitative theory of materials.
%\marginpar{Only nonrelativistic. What about SO?}
%, and I propose forms for the energy density and stress fields based upon the different roles of energy and stress in quantum theory of matter.

In \refsec{sec:mb-system-general} are basic definitions and in \refsecs{sec:ke-density} and \ref{sec:pe-density} 
are derivations that underlie the rest of the paper.  As shown in \refsec{sec:ke-density}, the kinetic energy density can be divided into a sum of two parts:
one part that takes into account exchange and correlation in a system of many interacting particles, which is shown to be a unique function defined at every point $\vecr$ independent of the choice of kinetic energy operators, and the rest which is not unique but is shown to be a function of only the density $n(\vecr)$. Thus the long-standing
controversies and proposals for choices for the kinetic energy density are relegated to terms which involve only the density. These terms are analogous to a single-particle problem with wave function $s(\vecr) \propto \sqrt{n(\vecr)}$, 
and various choices for the kinetic operators
can be elucidated by consideration of simple, exactly soluble examples. The consequences of interactions, discussed in \refsec{sec:pe-density}, are similar but there is the added issue that the energy can be associated with with potentials acting on particles or the field between particles. 
%\marginpar{Moved stress to after energy, May 16} 

The explicit form for the energy density $e(\vecr)$ is derived in \refsec{sec:e-density-dft} based on the role of the energy as the quantity that determines the ground state from the variational principle.   
For this purpose, the kinetic energy operator is $-\frac{1}{2} s\nabla^2 s$, and the interactions are included as potentials acting on the particles
due to external effects and interactions with other particles.  This form used in much of the literature\cite{che92,hammer95,rapcewicz98,MHCohen00,ernzerhof02,martin04-20,anderson-JCPA114-8884,yu2011,fredrickson2012} and  it naturally leads to density functional theory (DFT) with the interpretation as an equilibration of the energy density.  

As discussed in \refsec{sec:stress-field},
%and \refapp{app:stress-derivations}, 
a stress field  $\sigma_{\alpha \beta}(\vecr)$ is a property of the ground state that is related to forces.
It can also be divided into a part that includes effects of exchange and correlation, which is well-defined, plus a part that depends only on the density. 
For the latter terms,  the kinetic contribution to $\sigma_{\alpha \beta}(\vecr)$ can be correctly described by only one
combination of the kinetic energy operators $\frac{1}{2}[s\nabla^2 s - |\nabla s|^2]$, which was derived in the classic  works of
\Schr\ \cite{schr27}, Fock\cite{fock30}, and Pauli\cite{pauli33}, and in more recent work\cite{RMM-nielsen85,godfrey88,rogers02}  In the present paper, additional aspects are brought out 
%in \refsec{sec:stress-field} 
by application to simple models in \refapp{app:stress-square-wells}.  
%It turns out that this reveals interesting issues about kinetic energy in the classically forbidden region at the level of an undergraduate physics course.  
For Coulomb interactions stress is properly described in the Maxwell form in terms of electric fields, which are related to forces, whereas potentials are not directly relevant.
%\cite{RMM-nielsen85,godfrey88,rogers02}. 
 
% \refSec{sec:more} includes some practical considerations and the 

%\marginpar{July25 - Interpretation in \refsec{sec:relation} is new in this version.}

\refSec{sec:relation} provides a resolution of apparent contradictions between expressions for anergy and stress densities. The definition of a local energy in terms of the pressure in \refeq{eq:energy-pressure}, often attributed to Pauli, Schrodinger, and others, is indeed correct for a homogeneous system, but the interpretation as a local energy is 
% a local approximation analogous to the Thomas-Fermi approximation. 
an approximation.  This is not an ambiguity.  The correct results are found by consistently recognizing the differences between the stress field and the energy density.  Each is well defined and each is useful for a specific purpose.

It is important to emphasize that many quantities are independent of the issues addressed in this paper.
The total energy and macroscopic stress are integrated quantities that are
well-defined by integrals over a unit cell.  Surface energy and stress are defined
by integrals over the surface region\cite{che92,che92a,rapcewicz98,filippetti00}. An integral over the region around an atom specified by the condition $\nabla n(\vecr) = 0$ determines a unique kinetic energy, which is the basis for the ``Atoms in Molecules'' interpretation  \cite{bader90,bader91}.  For such cases, the present analysis may be useful
for understanding and finding efficient methods, but all choices must lead to the same result.

\section{Many interacting particles in an external potential}
\label{sec:mb-system-general}

The nonrelativistic Hamiltonian for particles of mass $m$ in an external potential
$V_{ext}(\vecr)$ can be written as
\begin{equation}
 {\hat{H}} = - \frac{\hbar^{2}}{2m} \sum_{i}{\nabla}^{2}_{i}
 + \sum_{i} V_{ext}(\vecr_i)
 + \frac{1}{2}\sum_{i\neq j} V_{int}(\modrij),
\label{eq:gen-ham}
\end{equation}
where we do not consider magnetic fields or relativistic effects. 
Much of the analysis is valid for arbitrary systems but we focus on electrons in materials where the nuclei are considered to be static and define the external potential.\footnote{For Coulomb interactions in condensed matter, one must take care to ensure the system is neutral and the sums are treated properly to specify the standard state of the material where the average of the Coulomb potential is defined to be zero, in which case the potential is fixed by the charge density in the bulk, as explained in \cite{martin04-20}  
}   
The wave function $\Psi(\vecr_1,\vecr_2,\ldots,\vecr_N)$ is a function of the coordinates
of all the particles $\vecr_1,\vecr_2,\ldots,\vecr_N$ and  is the solution of the many-body \Schr\ equation
\begin{equation}
         {\hat{H}} \Psi = E \Psi
\label{eq:schr-eq}
\end{equation}
and is here defined to be normalized:
\begin{equation}
        1 = \int d\vecr_1 d\vecr_2 \ldots d\vecr_N
        |\Psi(\vecr_1,\vecr_2,\ldots,\vecr_N)|^2.
\label{eq:gen-norm}
\end{equation}
Here we consider the stationary ground state which is the minimum energy solution of \refeq{eq:schr-eq}.
Spin is not indicated explicitly, but it can be included as a
composite variable along with the position $\vecr_i$ of each
particle $i$.

A density is defined by an expectation value; for example, the particle density is given by
\begin{equation}
        n(\vecr) =
\frac{\langle\Psi|{\hat{n}}(\vecr)|\Psi\rangle}{\langle\Psi|\Psi\rangle}
 = N \, \int d\vecr_2 \ldots d\vecr_N |\Psi(\vecr,\vecr_2,\ldots,\vecr_N)|^2,
\label{eq:gen-rho}
\end{equation}
where ${\hat{n}}(\vecr) = \sum_{i=1,N} \delta(\vecr - \vecri)$
is the  density operator. The symmetry of the \wf\ has been used to write the
integral in terms of only the first argument of $\Psi$ and
multiplied by the factor of $N$.
 The density \refeq{eq:gen-rho}
satisfies the integral condition:
\begin{equation}
N =  \int d\vecr n(\vecr),
\label{eq:n-density}
\end{equation}
We will consider the ground state in which case the particle density
is positive definite $n(\vecr) > 0 $ with no zeros.

The energy is the expectation value of the Hamiltonian in
\refeq{eq:gen-ham},
\begin{eqnarray}
        E_{tot} & = &
\int d\vecr \int d\vecr_2 \ldots d\vecr_N \nonumber \\
& &  \Psi^*(\vecr,\vecr_2,\ldots,\vecr_N)  \hat{H}
\Psi(\vecr,\vecr_2,\ldots,\vecr_N),
\label{eq:gen-E-tot}
\end{eqnarray}
 which can be expressed in terms the various contributions,
\begin{equation}
        E_{tot} =  \langle \hat{T} \rangle
          + \int d\vecr n(\vecr)V_{ext}(\vecr)
          + \langle \hat{V}_{int} \rangle,
\label{eq:gen-e-tot}
\end{equation}
where $\langle \hat{T} \rangle \equiv T$ is the expectation value
of the kinetic energy operator in the state $\Psi$  and $\langle \hat{V}_{int}
\rangle$ is the expectation value of the interaction energy.

In analogy to the particle density in \refeqs{eq:gen-rho} and \ref{eq:n-density}, the energy density
can be expressed in the form
\begin{equation}
E_{tot} =  \int d\vecr e(\vecr) = \int d\vecr \epsilon(\vecr) n(\vecr),
\label{eq:e-density}
\end{equation}
where the energy density $e(\vecr)$ is denoted by a lower case roman letter, and
the corresponding contributions can be defined by
\begin{equation}
e(\vecr) = t(\vecr) + n(\vecr)V_{ext}(\vecr) + e^V_{int}(\vecr).
\label{eq:total-e-density}
\end{equation}
The energy per particle $e(\vecr)/n(\vecr)$ is denoted by greek letters
\begin{equation}
\epsilon(\vecr) = \tau(\vecr) + V_{ext}(\vecr) + \epsilon^V_{int}(\vecr).
\label{eq:total-epsilon-density}
\end{equation}

For the stress field we follow conventional notation, \eg\ in \cite{RMM-nielsen85},
where the macroscopic average stress is denoted $\sigma_{\alpha \beta}$, and the
stress field is denoted $\sigma_{\alpha \beta}(\vecr)$, which is also divided into kinetic and interaction terms. Because of the nature of stress
as a variation of space, it is not useful to define stress density per particle at a point $\vecr$.
%\marginpar{Omit last sentence?}

\section{Kinetic energy density}
\label{sec:ke-density}

The total kinetic energy can be expressed as
\begin{equation}
 T=\int d \vecr t(\vecr)   
 \label{eq:gen-T-density}
\end{equation}
where the kinetic energy density can be expressed as
\begin{equation}
 t(\vecr) = - \frac{N}{2} \int d\vecr_2 \ldots d\vecr_N 
\Psi^*(\vecr,\vecr_2,\ldots\vecr_N)  \nabla_{\vecr}^2
\Psi(\vecr,\vecr_2,\ldots\vecr_N),  
\label{eq:gen-T-1}
\end{equation}
or
\begin{equation}
  t(\vecr) = \frac{N}{2} \int d\vecr_2 \ldots d\vecr_N    |\nabla_{\vecr}\Psi(\vecr,\vecr_2,\ldots,\vecr_N)|^2,
\label{eq:gen-T-2}
\end{equation}
%
%*** omitted text 5-1}
%
or any linear combination of the forms in \refeqs{eq:gen-T-1} and \ref{eq:gen-T-2}. For any of these forms the kinetic energy density per particle can be written as $\tau(\vecr) = t(\vecr)/n(\vecr)$,
%
% Either form can be regarded as a kinetic energy
% density $t(\vecr)= n(\vecr) \tau(\vecr)$.  
and the integral is the
total kinetic energy $T$, which follows from integration by
parts since boundary terms vanish for bound states where the density vanishes at the boundary or the boundary terms cancel in
the case of periodic boundary conditions.

%Omitted former Eq 14  following referee comment

The central result in this section is that the kinetic energy density
can be expressed as a sum of two terms, one that is a function only of the density $n(\vecr)$ and the other that takes into account all effects of exchange and correlation,
\begin{equation}
t(\vecr) = t_{n}(\vecr) + t_{xc}(\vecr),
\label{eq:t-B-F}
\end{equation}
where $t_{xc}(\vecr)$ is a unique, well-defined function no matter which form of the kinetic energy is used.
%This is a generalization of the approach of Savin\cite{savin95} and Levy and Gorling\cite{levy95} who considered independent fermions and the added kinetic energy due to exchange (see \refsec{sec:e-density-dft}).
%\marginpar{July25 - Added refs for EEF - important but no need to add discussion}
The two terms in \refeq{eq:t-B-F} can be derived by using the ``exact factorization'' approach \cite{hunter1975-probability,hunter1986-Exact,abedi-maitre-gross-PhysRevLett.105.123002,gonze-reining-2018,kocak-schild2021,kocak-schild-PhysRevResearch.5.013016}
to express the wave function s as a product
of a symmetric part $S$ and an antisymmetric part $\Phi$,
\begin{equation}
\Psi(\{\vecr\}) = S(\{\vecr\}) \Phi(\{\vecr\}).
\label{eq:Psi-S-Phi}
\end{equation}
where $\{\vecr\} = \vecr_1, \vecr_2 \ldots \vecr_N$.
We choose a simple form for the symmetric part: a product
wave function for $N$ non-interacting particles with no symmetry (sometimes called ``boltzmannons'') with the ground state
%density $n(\vecr)$, so that
%
\begin{equation}
S(\{\vecr\}) = s(\vecr_1) s(\vecr_2) \ldots s(\vecr_N),
\label{eq:S-def}
\end{equation}
where
\begin{equation}
 s( \vecr) = \left(\frac{n(\vecr)}{N}\right)^{1/2},
\label{eq:s-n-def}
\end{equation}
so that $s^2(\vecr)$ is proportional to the density and is
normalized to $1$. Inserting \refeq{eq:Psi-S-Phi} into the
expression \refeq{eq:gen-rho} for the density leads to
\begin{equation}
        n(\vecr)  = s^2(\vecr)\, N \, \int d\vecr_2 \ldots d\vecr_N s^2(\vecr_2) \ldots s^2(\vecr_N)
        |\Phi(\vecr,\vecr_2,\ldots,\vecr_N)|^2.
\label{eq:Phi-n}
\end{equation}
From the defintion of $s$ in  \refeq{eq:s-n-def}, it follows that
\begin{equation}
        1 = \int d\vecr_2 \ldots d\vecr_N s^2(\vecr_2) \ldots s^2(\vecr_N)
        |\Phi(\vecr,\vecr_2,\ldots,\vecr_N)|^2,
\label{eq:Phi-norm}
\end{equation}
at each point $\vecr$.  Taking derivatives of \refeq{eq:Phi-norm}
with respect to $\vecr$ we find for the first derivative,
\begin{eqnarray}
0  & = & 2 Re \int d\vecr_2 \ldots d\vecr_N s^2(\vecr_2) \ldots
        s^2(\vecr_N) \nonumber \\
 & & [ \; \Phi^*(\vecr,\vecr_2,\ldots,\vecr_N)
\nabla_{\vecr}\Phi(\vecr,\vecr_2,\ldots,\vecr_N)  \; ],
\label{eq:Phi-grad}
\end{eqnarray}
and for the second derivative,
\begin{eqnarray}
0  & = & 2 Re \int d\vecr_2 \ldots d\vecr_N s^2(\vecr_2) \ldots
        s^2(\vecr_N) \nonumber \\
 & & [ \; \Phi^*(\vecr,\vecr_2,\ldots,\vecr_N)
\nabla_{\vecr}^2\Phi(\vecr,\vecr_2,\ldots,\vecr_N) \nonumber \\
 & & \;\;\; + |\nabla_{\vecr} \Phi(\vecr,\vecr_2,\ldots,\vecr_N)|^2 \; ],
\label{eq:Phi-grad-sq}
\end{eqnarray}
which must be satisfied at each point $\vecr$.

The two expressions for the kinetic energy,
\refeq{eq:gen-T-1} and \refeq{eq:gen-T-2}, can be written as
(omitting the arguments $\vecr, \vecr_2, \vecr_3, \ldots
\vecr_N$ of each function and defining $\nabla_{\vecr} = \nabla$)
\begin{equation}
-S\Phi^* \nabla^2 (S\Phi) =  -(S \nabla^2 S) |\Phi|^2  - S^2
\Phi^*\nabla^2 \Phi - 2 S \nabla S \Phi^* \nabla \Phi,
\label{eq:S-Phi-integrand-1}
\end{equation}
and
\begin{equation}
|\nabla (S\Phi)|^2 = (\nabla S)^2 |\Phi|^2 + S^2 |\nabla \Phi|^2
+ 2 S \nabla S \Phi^* \nabla \Phi.
\label{eq:S-Phi-integrand-2}
\end{equation}
The kinetic energy density at each point $\vecr$ is defined by the integral over $\vecr_2, \vecr_3, \ldots \vecr_N$.
The integral over the last term in each equation vanishes because it $s(\vecr) \nabla_{\vecr} s(\vecr)$ multiplied by the integral on the right hand side of \refeq{eq:Phi-grad} which vanishes.
The first and second terms (multiplied by one-half) are respectively the two contributions to the kinetic energy density $t_{n}(\vecr)$ and  $t_{xc}(\vecr)$ in \refeq{eq:t-B-F}  which are analyzed in the following subsections.

\subsection{Exchange-correlation kinetic energy density $t_{xc}(\vecr)$}
\label{subsec:ke-xc}

The middle terms on the right
hand sides of \refeq{eq:S-Phi-integrand-1} and
\ref{eq:S-Phi-integrand-2} are two
representations of the exchange-correlation part of the kinetic
energy density $t_{xc}(\vecr)$ in \refeq{eq:t-B-F}.  Using
\refeq{eq:Phi-grad-sq} and integrating over $\vecr_2,\ldots \vecr_N$,
it follows immediately that these two forms are equal at each
point $\vecr$. Thus the exchange-correlation part of the kinetic
energy density is a unique function of position $\vecr$,
$t_{xc}(\vecr) = n(\vecr) \tau_{xc}(\vecr)$, with
$\tau_{xc}(\vecr)$ given by
\begin{eqnarray}
\label{eq:t-xc}
& & \tau_{xc}(\vecr) = \nonumber \\
& - &  \frac{1}{2}
 \int d\vecr_2 \ldots d\vecr_N s^2(\vecr_2) \ldots s^2(\vecr_N)
\nonumber \\
& & [ \Phi^*(\vecr,\vecr_2,\ldots,\vecr_N) \nabla_{\vecr}^2
\Phi(\vecr,\vecr_2,\ldots,\vecr_N)] \\
&=&  \frac{1}{2}  \int d\vecr_2 \ldots d\vecr_N s^2(\vecr_2)\ldots
s^2(\vecr_N) |\nabla_{\vecr} \Phi(\vecr,\vecr_2,\ldots,\vecr_N)|^2
\nonumber,
\end{eqnarray}
where we have used the relation $N s^2(\vecr) = n(\vecr)$.
Note that the exchange-correlation kinetic energy density $\tau_{xc}(\vecr) $ is always positive, as shown explicitly by the form in the second line of \refeq{eq:t-xc}. Thus the present formulation embodies the fact that, in the ground state of any  system of interacting, correlated particles, the kinetic energy is increased at every point compared to a system of uncorrelated ``boltzmannons'' at the same density $\tau_n(\vecr)$.

%\marginpar{Added sentences after \refeq{eq:t-xc} & moved next par. May 20}

The result that the kinetic energy due to exchange-correlation is uniquely determined at each point can also be understood independently of the mathematical derivation.  The excess kinetic energy due to exchange and correlation is determined by the relative motion of particles, and the effects extend only to distances within which the particles are correlated and/or are affected by the exclusion principle.  This is analogous to the exchange-correlation interaction energy, which depends on the relative positions of the particles and is expressed in \refeq{eq:xc-epsilon-av-nxc} below as the exchange-correlation hole, which goes to zero at large distance from point $\vecr$.
%To the knowledge of the author there is not a way to measure the corresponding kinetic correlations, but they are expected to extend only over distance similar to the exchange-correlation hole and vanish at large relative distance.
Thus $\tau_{xc}(\vecr)$ is an example of kinetic energy over a region of space around point $\vecr$ with zero boundary conditions. Just as in other cases with zero boundary conditions, integration by parts must lead to the same result for either form of the kinetic energy. Thus, even though it is difficult to calculate for correlated particles, it is uniquely defined at any point $\vecr$ independent of the form chosen for the kinetic energy.
For independent particles, the expressions are straightforward, as described in \refsec{subsec:ind-part-tau-x} below.

\subsection{Density-dependent kinetic energy density $\tau_n(\vecr)$}
\label{subsec:tau-n}

 The contribution to the kinetic energy density of the first terms on the right hand sides of 
 \refeqs{eq:S-Phi-integrand-1} and  \ref{eq:S-Phi-integrand-2}
 respectively involve only the function
 $s(\vecr) = \sqrt{n(\vecr)/N}$ and its
derivatives, since the integral of $S^2|\Phi|^2$ over $\vecr_2,
\ldots \vecr_N$ is unity at every point $\vecr$, as required in
\refeq{eq:Phi-norm}. The result is the same as for  $N$ non-interacting ``boltzmannons''
(particles with simple product \wf) with density $n(\vecr)$, \ie\ with each particle having the same
wave function\ $s(\vecr)$.  The resulting kinetic energy density per particle is not
unique and may be chosen to be any linear combination of the two forms
\begin{equation}
\tau_{n}(\vecr) =  \left[A \; \frac{1}{2} | \nabla s(\vecr)|^2
+ (1-A) \; \left( -\frac{1}{2} s(\vecr) \nabla^2 s(\vecr) \right) \right]
\label{eq:t-boltzmannon},
\end{equation}
where $A$ is arbitrary.  \refEq{eq:t-boltzmannon} can be expressed in terms of the density as
\begin{equation}
\tau_{n}(\vecr) = \frac{1}{8} \frac{|\nabla n(\vecr)|^2}{n(\vecr)}
 + (1-A) \; \left[ -\frac{1}{4} \nabla^2 n(\vecr) \right]
\label{eq:t-boltzmannon-density},
\end{equation}
and all of the possible expressions for $\tau_{n}(\vecr)$ are related by addition of a term involving only the Lapacian of the density $\nabla^2 n(\vecr)$.

Thus the issue of nonuniqueness of the
kinetic energy density has been reduced to the simplest possible
form involving only the density and all the issues are the same as
in a one-particle problem for which the wave function is $s(\vecr)$.
A definitive form for $t_{n}(\vecr)$
can be specified {\it only} if we add further requirements. 
%The well-known Weizsacker form\cite{weiz35} for the first order correction to the Thomas-Fermi approximation  corresponds to $A=1$. 
As derived in \refsec{sec:e-density-dft}, the correct choice for minimizing the energy is $A=0$, whereas in  \refsec{sec:stress-field} 
for stress the value is $A=1/2$.

A comment may be in order to avoid confusion.  The density-dependent $\tau_{n}(\vecr)$ has nothing to do with density-dependent approximations, such as the Thomas-Fermi or orbital-free approximations (see \refsec{subsec:useful}), which are approximations for the full kinetic energy $\tau(\vecr)$. For example, for the homogeneous electron gas the density $n(\vecr)$ is constant and $\tau_{n}(\vecr) = 0$, whereas the Thomas-Fermi approximation is a local approximation in terms of the kinetic energy $\tau^{ip}$  for many independent Fermions with constant density $n$.

\subsection{Expressions for independent electrons}
\label{subsec:ind-part-tau-x}

Since the expressions for $\tau_{xc}(\vecr)$ in \refeq{eq:t-xc} are rather formal and hard to appreciate, it is useful
give the expressions for kinetic energy density $\tau^{ip}(\vecr)$ for independent electrons (or other fermions) where the wave function can be written in terms of single-particle orbitals $\psi_i(\vecr) = s(\vecr) \phi_i(\vecr)$.  Then $\tau^{ip}(\vecr)$ 
%in \refeq{eq:t-B-F} 
is the sum of $\tau_n(\vecr)$, which is unchanged since it depends only on the density, and the rest which can be written as
\begin{equation}
\tau^{ip}_{x}(\vecr) = -\frac{1}{2N} \sum_i \phi_i^*(\vecr) \nabla^2 \phi_i(\vecr)
= \frac{1}{2N} \sum_i |\nabla \phi_i(\vecr)|^2.
\label{eq:t-x-only-phi}
\end{equation}
However, for independent particles it is most straightforward to calculate $\tau^{ip}_{x}(\vecr)$  as the difference  $\tau^{ip}_{x}(\vecr) = \tau^{ip}(\vecr) - \tau_n(\vecr)$. Although both $\tau^{ip}$ and $\tau_n$ depend on the choice of forms for the kinetic energy, the difference is invariant:  
\begin{equation}
\tau^{ip}_{x}(\vecr)
= \frac{1}{2N} \sum_i |\nabla \psi_i(\vecr)|^2 
- \frac{1}{2} |\nabla s(\vecr)|^2,
\label{eq:t-x-only-psi-1}
\end{equation}
or equivalently
\begin{equation}
\tau^{ip}_{x}(\vecr)
= -\frac{1}{2N} \sum_i \psi_i(\vecr) \nabla^2 \psi_i(\vecr) 
+ \frac{1}{2} s(\vecr) \nabla^2 s(\vecr).
\label{eq:t-x-only-psi-2} 
\end{equation}
 Here $\tau^{ip}_{x}(\vecr)$ is termed the ``exchange kinetic energy density''  since it is $\tau^{ip}_{xc}(\vecr)$ in cases where there is no correlation; however, it is often called the Pauli kinetic energy\cite{march1986} since it is due to the exclusion principle.
 
The kinetic energy 
% due to the fermion character of the particles is contained in $\tau^{ip}_{x}(\vecr)$, and it
has a well-defined physical meaning, the curvature of the exchange hole\cite{stoll80,becke83,luken84,bader96} at point
$\vecr$, which can be shown to be the relative kinetic energy of
pairs of electrons\cite{dobson91} independent of the form of the kinetic energy.  These conclusions  are well-established, as derived, \eg in \cite{stoll80,becke83,luken84,bader96} and in appendix H of \cite{martin04-20}. For example, the electron localization function (ELF)\cite{becke90,savin-doi:10.1002/anie.199718081} 
and meta-generalized-gradient-approximation (meta-GGA) functionals\cite{dun84,becke89,perdew99,ernzerhof02,perdew-RMM:QUA25100}  
% are formulated in terms of \refeq{eq:t-x-only-psi-1} or \ref{eq:t-x-only-psi-2} 
% These forms are 
as discussed further in \refsec{sec:KS} on the Kohn-Sham approach.

%Dec 5 - commented out - \marginpar{CHECK refs to Savin, Levy, Gorling}

\section{Interaction energy}
\label{sec:pe-density}

 The interaction energy involves an additional consideration since the energy can be associated with the particles or with the interaction treated as a field. In a nonrelativistic Hamiltonian such as in \refeq{eq:gen-ham}, the formulation is in terms of particles and potentials, and the potential energy is given by the last two terms in \refeqs{eq:gen-e-tot} - \ref{eq:total-epsilon-density}.
 In this case one can define a potential energy density that is analogous to the kinetic energy density as a sum of two terms: one that involves only the density, which is not unique, and another that incorporates the effects of exchange and correlation among the particles, which is uniquely defined at each point $\vecr$ in terms of physically measurable quantities.

Since the external potential acts locally on particles at each point $\vecr$, the contribution to the energy density per particle is simply $V_{ext}(\vecr)$.
For the energy due to interactions, it is most useful to express the last term as the sum of a mean-field Hartree term and the remainder which is due to exchange and correlation among the particles,
\begin{equation}
\langle \hat{V}_{int} \rangle = E_{Hart} + E_{xc}^V,
\label{eq:E-V-total}
\end{equation}
where
\begin{eqnarray}
        E_{Hart} &=&   \frac{1}{2} \int d^3\vecr d^3\rp
        n(\vecr) V_{int}(|\vecr - \vecrp|)n(\vecrp)\\
        &\equiv& \frac{1}{2} \int d^3\vecr n(\vecr) V_{Hart}(\vecr).
\label{eq:E-Hart}
\end{eqnarray}
The Hartree term is the interaction of a continuous density with itself; for a given form of the interaction $V_{int}(|\vecr - \vecrp|)$, $E_{Hart}$ is a function only of the density.
With these definitions, the potential energy density per particle can be written as
\begin{equation}
\epsilon^V_{int}(\vecr) = \epsilon^V_{n}(\vecr) + \epsilon^V_{xc}(\vecr),
\label{eq:V-epsilon-density}
\end{equation}
where $\epsilon_{n}(\vecr) = V_{ext}(\vecr) + \frac{1}{2} V_{Hart}(\vecr)$ is the sum of potential energy terms that are a function only of the density $n(\vecr)$ and  $\epsilon^V_{xc}(\vecr)$ is energy density per particle due to exchange and correlation among the particles.

%The notation is chosen to emphasize to  the close analogy with the possible expressions for the ``boltzmannon'' part of the kinetic energy in \refeqs{eq:t-boltzmannon} and \ref{eq:tau-boltzmannon}.

%\subsubsection{Potential energy due to exchange and correlation}

The exchange-correlation energy $E_{xc}^V$ contains the effects due to the fact that the
system is made up of interacting particles, which obey the exclusion principle in the case of fermions.
Even though it may be very difficult to calculate, the
important point for our purposes is that the contribution to the potential energy density
$\epsilon_{xc}^V(\vecr)$ can be specified at every point $\vecr$ as the potential energy per particle at point $\vecr$ due to exchange and correlation among
the particles, which is the energy of interaction of the electron at point $\vecr$ with its
exchange-correlation hole,
\begin{equation}
        \epsilon_{xc}^V(\vecr) =   \frac{1}{2} \int d^3\rp
        V_{int}(|\vecr - \vecrp|) n_{xc}(\vecr,\vecrp) .
\label{eq:xc-epsilon-av-nxc}
\end{equation}
This is a well-defined function since  $n_{xc}(\vecr,\vecrp)$ is the physically measurable
correlation function among the particles.
%and the factor of $1/2$  requirement due to the indistinguishability of electrons.
Note that $\epsilon_{xc}^V(\vecr)$ is different from the Kohn-Sham definition $\epsilon^{KS}_{xc}(\vecr)$. Whereas $\epsilon_{xc}^V(\vecr)$ is given directly in terms
of the physical correlation function in \refeq{eq:xc-epsilon-av-nxc},
% or  \ref{eq:xc-epsilon-av-hole},
in the Kohn-Sham formulation $\epsilon^{KS}_{xc}(\vecr)$ includes all effects of correlation including a kinetic contribution, as discussed in \refsec{sec:e-density-dft}.

However, this is not the only possible choice.
The interaction terms could be associated spatially with the interactions between the particles instead of potentials acting on the particles. For Coulomb interactions this is
the Maxwell form in terms of the electric field\cite{lan60,jac62}.  
In that case the interaction energy can also be divided into an average mean-field term, which depends only on the density, and a part that is due to correlation of the particles.  
This is discussed further in \refsec{sec:stress-field} since it is directly relevant for the stress field.

\section{Energy density and Density Functional Theory}
\label{sec:e-density-dft}

Energy plays a special role in the theory.   
Minimization of the energy determines the ground state whereas other properties, such as the stress in \refsec{sec:stress-field}, are determined by the ground state. In the original many-body problem, this is expressed as the minimization of the energy is \refeqs{eq:gen-E-tot} and \ref{eq:gen-e-tot}  with respect to the many-body wave function $\Psi(\vecr,\vecr_2,\ldots,\vecr_N)$. The  energy density formulation 
%in terms of the density in \refeqs{eq:total-e-density} and \ref{eq:total-epsilon-density}, which 
provides an alternative approach by breaking the problem into two parts, one part a function only of the density and the other part taking into account all effects of exchange and correlation.  The topic of this section is the resolution of the choices for the form of the energy density and the demonstration that it is a way to derive density functional theory with an interpretation as the equilibration of the energy density.

%\marginpar{New section May 20}
\subsection{Variational principle and minimization of fluctuations}
\label{subsec:variational}

The analysis of \refsecs{sec:ke-density} and \ref{sec:pe-density} has shown that for a system of many interacting particles in an external potential, the ground state energy density is the sum of kinetic and interaction terms, each of which can be expressed as a sum of mean-field plus exchange-correlation terms as given in \refeqs{eq:t-B-F} and \ref{eq:V-epsilon-density}. The exchange-correlation contribution to the kinetic energy density per particle $\tau_{xc}(\vecr)$ is a unique function, the same no matter which form of the kinetic operators is used. However, the contribution $\tau_{n}(\vecr)$ is not unique and may be expressed in terms of   $-\frac{1}{2}s\nabla^2 s$,  $\frac{1}{2}|\nabla s|^2$, or a linear combination as expressed in
\refeqs{eq:t-boltzmannon}
and \ref{eq:t-boltzmannon-density}.
The interaction energy energy density can be constructed in two ways: as potentials acting on particles or as the field between particles. In each case the contributions 
% $\epsilon^V_{n}(\vecr)$ and $\epsilon^V_{xc}(\vecr)$ 
are well defined, but they are different and one can make either choice.

At this point there is no compelling reason to choose one form or another; the various forms are all allowed expectation values of the ground state wave function and all integrate to the total energy. 
%ZZZZZZZZZ
However, there is another principle that can be used, that fluctuations in the energy density $\epsilon(\vecr)$ vanish. 
The only choice that satisfies this criterion is the expression for kinetic energy 
%in \refeq{eq:gen-T-1} 
in terms of $-\frac{1}{2}\nabla^2s$ and the interaction energy in terms of potentials acting on particles, 
%in \refeq{eq:E-Hart} and \ref{eq:xc-epsilon-av-nxc}, 
\ie an energy density given by
\begin{eqnarray}
\label{eq:epsilon-DFT}
\epsilon(\vecr) &=& -\frac{1}{2}s(\vecr)\nabla^2s(\vecr) + \tau_{xc}(\vecr) \nonumber \\ &+& V_{ext}(\vecr) + \frac{1}{2} V_{Hart}(\vecr)  + \epsilon^V_{xc}(\vecr).
\end{eqnarray}
In  \refsecs{sec:ke-density} and \ref{sec:pe-density} it was emphasized that the kinetic energy for $s(\vecr)$ is like a single-body problem; however, minimization of the energy in terms of the energy density \refeq{eq:epsilon-DFT} is not the same as solving a single-body \Schr\ equation, since the expressions for $\tau_{xc}(\vecr)$ and $\epsilon^V_{xc}(\vecr)$ involve the density. Ways to minimize the energy are discussed in the following section and in \refsec{subsec:useful}.

\subsection{Density functional theory}
\label{subsec:dft}

The present approach provides a way to derive density functional theory in terms of the energy density.  
The difficult parts of the many-body problem are the exchange-correlation terms, $\tau_{xc}(\vecr) + \epsilon^V_{xc}(\vecr)$  which are expressed in terms of the many-body wave function in \refsecs{sec:ke-density} and \ref{sec:pe-density}. In the process of determining the ground state variationally, these contributions to the energy density must be determined for densities  $n(\vecr) = N s( \vecr)^2$, even when it is not the ground state density.  This is analogous to the formulation of DFT by Levy and Lieb (LL) \cite{levy85,lieb85}, where expressions are formulated in terms of the full many-body way function $\Psi$.  
However, the present derivations do not depend on the Levy-Lieb demonstration that the exchange-correlation terms are implicit functions of the density. 

One way to derive an analytic formulation of DFT is minimization of the energy density in \refeq{eq:epsilon-DFT}  
with the constraint that the particle number is fixed $\int n(\vecr) =N$.  Using the method of LaGrange multipliers as done by Savin\cite{savin95} and Levy and Gorling\cite{levy95}, the variational equation can be expressed as
\begin{equation}
\frac{\delta}{\delta n(\vecr)}\left[ E_{tot} -
\mu \left(\int n(\vecr) - N \right) \right] =
\frac{\delta E_{tot}}{\delta n(\vecr)} - \mu = 0,
 \label{eq:var-eq-density}
\end{equation}
where $\mu$ is the chemical potential and $E_{tot}$ is the total
energy given by \refeq{eq:gen-e-tot}.  The result is that $s(\vecr)$ is the ground state of the \Schr-like equation
\begin{equation}
-\frac{1}{2}\nabla^2 s(\vecr) + V_{eff}(\vecr)s(\vecr)
= \mu s(\vecr),
 \label{eq:schr-eq-density}
\end{equation}
where the effective potential is the variational derivative,
\begin{equation}
V_{eff}(\vecr) = V_{ext}(\vecr) + V_{Hart}(\vecr) +  \frac{\delta}{\delta n(\vecr)} \int d \vecrp \, \left[
 \tau_{xc}(\vecrp) + \epsilon^V_{xc}(\vecrp) \right],
 \label{eq:v-eff-eq-density}
\end{equation}
and $s$ is normalized $\int s(\vecr)^2 =1$.
Thus the problem of finding the ground state of the many-body system has been recast as the problem of equilibrating the energy density $\epsilon(\vecr)$ with the same chemical potential
at all points $\vecr$ in the system.  
% The result is equivalent to the well-known density functional theory; however, the present derivation does not invoke the Hohenberg-Kohn theorem that all properties of the system are in principle functionals of the density in the ground state. 
%  Instead, it is an alternative approach specifically for the ground state, with the demonstration that there is a well-defined energy density given explicitly
%  in \refeq{eq:epsilon-DFT}. The Hohenberg-Kohn theorem is another step not needed at this point. 

The equations for minimization are self-consistent equations involving the density and the effective potential $V_{eff}(\vecr)$ in \refeq{eq:v-eff-eq-density}.  The formal approach in density functional theory is that the potential is determined by the density $n(\vecr)$ which is found by minimization.  However, the more practical approach (for example, in the Kohn-Sham method) is that the equations determine the density given the potential $V_{eff}(\vecr)$.  In this sense the energy density $\epsilon(\vecr)$  in \refeq{eq:epsilon-DFT}
 is the basic quantity; in the minimization process, the density $n(\vecr)$ follows the direction determined by the gradient of $\epsilon(\vecr)$ with respect to the density $n(\vecr)$.

\subsection{The Kohn-Sham approach}
\label{sec:KS}

The Kohn-Sham construction of an auxiliary system of independent particles provides a practical way to use the formalism developed so far.  For the auxiliary system the energy density can be written as
\begin{equation}
\epsilon^{KS}(\vecr) = \tau^{ip}(\vecr) + V_{ext}(\vecr) + \frac{1}{2} V_{Hart}(\vecr)  + \epsilon^{KS}_{xc}(\vecr),
\label{eq:epsilon-KS}
\end{equation}
where $\tau^{ip}(\vecr)$ is the kinetic energy density for independent particles.  It can be expressed in the form of \refeq{eq:epsilon-DFT}  if we recognize that $\tau^{ip}(\vecr) = -\frac{1}{2}s(\vecr)\nabla^2s(\vecr) + \tau^{ip}_{x}(\vecr)$ and the Kohn-Sham functional,
\begin{equation}
 \epsilon^{KS}_{xc}(\vecr) = \tau_{xc}(\vecr) -\tau^{ip}_{x}(\vecr) + \epsilon^{V}_{xc}(\vecr),
\label{eq:epsilon-xc-KS}
\end{equation}
includes both the xc interaction term $ \epsilon^{V}_{xc}(\vecr)$ in \refeq{eq:xc-epsilon-av-nxc} (the difference between the interaction energy density in the correlated system and the mean-field term) and the xc kinetic energy term (the difference between the kinetic energy density in the correlated system and the mean-field independent-particle system).  

%\marginpar{June25 - Need to add idea  of Pauli potential.}

Practical calculations are feasible because the kinetic energy density $\tau^{ip}(\vecr)$  can be calculated readily, and the self-consistent procedure illustrates the idea that the energy density can be considered to be the fundamental quantity in the minimization process. Even though formally the kinetic energy is a functional of the density, in practice it is calculated from a \Schr-like equation where the wave functions are determined by the potential, \ie by the energy density.

The energy density formulation provides illuminating ways to understand aspects of functionals. For example, meta-GGA functionals\cite{dun84,becke89,perdew99,ernzerhof02,perdew-RMM:QUA25100}  involve a kinetic energy density $\tau^{ip}(\vecr)$ defined in terms of the Kohn-Sham orbitals. As discussed in \refsec{subsec:ind-part-tau-x} and in \cite{martin04-20,perdew-RMM:QUA25100}, it actually involves only the exchange part $\tau^{ip}_x(\vecr)$ which is a well defined function of $\vecr$.  This has the sensible requirement that it should involve only the fact that the particles are fermions. The forms of the functionals can be interpreted in terms of the short-range shape of the exchange hole, which is determined by $\tau^{ip}_x(\vecr)$.  
It would be interesting to investigate ways in which the nature of the correlation hole
could be related to the correlation kinetic energy $\tau_c(\vecr)$.

\subsection{Usefulness of the energy density}
\label{subsec:useful}

The previous discussion emphasized ways the energy density provides illuminating interpretations of density functional theory.
It may also lead to practical possibilities using the division of the problem into a part involving the density and parts involving exchange and correlation. For example, there are various very efficient algorithms for dealing with a single function $s(\vecr)$  and the long-range Coulomb potential which is a functional of the density.  The Car-Parrinello approach\cite{car85} could be used to define a classical equation for the density that can be solved using molecular dynamics for either actual dynamics or achieving the equilibrium state. The rest of the problem could be treated in various ways, \eg the full many-body problem using the Levy-Lieb formulation or the Kohn-Sham auxiliary systems using efficient methods for dealing with many orthonormal orbitals.

A possibility is to view the present formulation as construction of an auxiliary density, which equals the actual density only at the solution. This would be analogous to the Kohn-Sham auxiliary system in the sense that it is designed to calculate only the ground state density and energy, not other properties of the system.

%\marginpar{June25 - mention Pauli potential give refs here.}

There is a long history of approximate methods in terms of the density
starting with the Thomas-Fermi approximation developed in the 1920s, which is the local density approximation for $\tau_{x}(\vecr)$ from the homogeneous electron gas, and
improved approximations like the Weizsaker correction\cite{weiz35} for an inhomogeneous gas.
More recently ``orbital free'' approaches\cite{mi-acs.chemrev.2c00758,witt2018} are density functional methods in terms of only the density instead of the Kohn-Sham approach where the kinetic energy is expressed in terms of wave functions. Progress has been made to find useful functionals, e.g., for application to semiconductors\cite{huang-PhysRevB.81.045206,shao-PhysRevB.104.045118}. See also \cite{francisco-JPC-A-2024} and references therein for ``deorbitalization'' strategies. Although aspects of the present work are closely related to derivations in the orbital-free approach, this is not discussed further here since it is not the focus of this paper.

A different use of the density is as a post-calculation analysis tool. For this purpose
a convenient form is to rewrite the expression in terms of the eigenvalues of the Kohn-Sham equations
\begin{equation}
\epsilon(\vecr) =  \sum_i \epsilon_i|\psi_i(\vecr)|^2
- \frac{1}{2} V_{Hartree}(\vecr) + [\epsilon_{xc}(\vecr) -
V_{xc}(\vecr)],
 \label{eq:epsilon-weighted-eigenvalues}
\end{equation}
where the first term includes the kinetic energy and potential terms in the
Schr\"{o}dinger-like equations and the last two terms account for over-counting
in the eigenvalues. This has been termed the eigenvalue weighted density
in a paper entitled "Total energy density as an interpretative tool" \cite{MHCohen00} and used, \eg\ in  \cite{hammer95}. 
% "Local Chemical Reactivity of a Metal Alloy Surface"

\section{Stress fields}
\label{sec:stress-field}

%\marginpar{Sign of stress corrected May 21!}

Stress fields $\sigma_{\alpha \beta}(\vecr)$ are considered in this paper along with energy density in order to clarify their similarities and differences. In much of the literature a local pressure, the trace of a local stress tensor, is used to define an energy density that is different from what has been derived in the previous sections.  Indeed, the review by Anderson and coworkers\cite{anderson-JCPA114-8884} concludes
``The ambiguity in the local kinetic energy can be considered to be inherited from the ambiguity in the local electronic stress tensor.'' One of the major goals of the present work is to clarify the difference between energy and stress, which leads unambiguously to differences between an energy density and a stress field, each useful in well-defined way. 

\subsection{Stress and Force}
\label{sec:stress-field-force}

The average stress $\sigma_{\alpha \beta}$ of a macroscopic system is
the derivative of the total energy per unit volume with respect to the strain tensor.
%$\epsilon_{\alpha \beta}$.
%It has the same units as energy density but it is fundamentally different.
Expressions for quantum systems have been given by Nielsen and Martin (NM)
\cite{RMM-nielsen85} based upon a uniform scaling of space, which are a generalization of the well-known virial theorem for the pressure. 
% See that paper for 
% references to early work starting in the 1920's.
Like the total energy, the expressions are well defined independently of the choice of kinetic energy operators and whether the interactions are described in terms of Coulomb potentials or electric fields. 

However, difficulties arise when one attempts to  construct a local stress field 
$\sigma_{\alpha \beta}(\vecr)$ at each point $\vecr$ on a microscopic scale. 
A stress field is defined by the condition that its divergence is the force field $\vecf(\vecr)$ acting on particles at point $\vecr$,
\begin{equation}
\sum_{\beta} \nabla_{\beta} \sigma_{\alpha \beta}(\vecr) =
f_{\alpha}(\vecr) = - n(\vecr) \frac{V_{ext}(\vecr)}{d \vecr_{\alpha}}.
\label{eq:gen-stress-force}
\end{equation}
The scalar local pressure is 
\begin{equation}
p(\vecr) = - \frac{1}{d} \sum_{\alpha}  \sigma_{\alpha \alpha}(\vecr). 
\label{eq:gen-stress-pressure}
\end{equation}
where $d$ is the dimension. 
In one dimension the stress is determined by \refeq{eq:gen-stress-force} except for a constant.  However, for space dimension $d > 1$, \refeq{eq:gen-stress-force} does not uniquely determine all components of the tensor $\sigma_{\alpha \beta}(\vecr)$,
since the curl of any vector field can be added with no change in the forces. This is a fundamental issue of nonuniqueness in both 
classical\cite{irving50,lan59,schofield82,wajnryb95} and quantum\cite{RMM-nielsen85,godfrey88,filippetti00,rogers02,martin04-20,anderson-JCPA114-8884}
theories.

An alternate derivation of the stress field in terms of a metric tensor field by Rogers and Rappe\cite{rogers02} is a local scaling of space analogous to the uniform scaling used to derive the macroscopic stress\cite{RMM-nielsen85}. The results are equivalent with the same issue of nonuniqueness due to addition of a field with zero divergence. 

% In this paper, we show that it is sufficient to consider stress and force in one dimension, where there is no ambiguity, to derive conclusively the form of the kinetic energy terms, which are different from those for energy in the previous sections, and to understand the physical meaning.  Simple examples in higher dimensions provide strong arguments for the choice of the curl terms in higher dimensions.     

\subsection{Stress field for the electronic system}
\label{sec:stress-field-electrons}

The formulation of the stress field for the electronic system takes advantage of the fact that the system is in the ground state, which is determined by the \Schr\ equation. 
Because there are subtle issues in constructing a stress field, we give a short summary of the major points. 
%in \refsec{sec:stress-kinetic} to \refsec{sec:relation}.
\begin{itemize}
    \item 
    Even though energy density and stress are very different, much of the analysis in \refsecs{sec:ke-density} to \ref{sec:e-density-dft} applies with appropriate modifications.  As described in \refsecs{sec:stress-kinetic} and \ref{sec:stress-interactions} the stress field can be divided in to contributions from kinetic terms and interactions,
\begin{equation}
\sigma_{\alpha \beta}(\vecr)   = \sigma^t_{\alpha
\beta}(\vecr)  + \sigma^{int}_{\alpha \beta}(\vecr),
 \label{eq:stress-field-tau-V}
\end{equation}
where the exchange-correlation terms are uniquely defined. 

\item 
All issues of nonuniqueness are in terms of the density and are analogous to a single-body problem, which has been studied before in great detail, with the result given in \refeq{eq:stress-field-tau-s-text}. However, there is the problem of addition of an arbitrary field, the second line of \refeq{eq:stress-field-tau-s-text}, which is the source of ambiguity \cite{anderson-JCPA114-8884}.

\item 
Here we propose a way to determine the form of the kinetic operators with no ambiguity by first considering one-dimension and generalizing to higher dimension. The analysis in \refsec{sec:stress-simple} shows clearly the reasons for the form of the kinetic stress and the difference from the kinetic energy.

\item 
Although there is no rigorous proof, \refsec{sec:stress-2-3d} we give strong arguments that for this problem the only reasonable choice is $B=0$, and the stress given by the first line of \refeq{eq:stress-field-tau-s-text}.

\item 
As described in \refsec{sec:stress-interactions}, interactions are properly treated as force fields, \eg in terms of electric fields for Coulomb interactions, unlike the results in \refsecs{sec:pe-density} and \ref{sec:e-density-dft} for the energy density.  Issues for Kohn-Sham calculations are discussed in \refsec{sec:stress-KS}. 

\item 
The results are brought together in \refsec{sec:relation}, where we clarify the differences between the local pressure $p(\vecr)$ and energy density $e(\vecr)$ and the consequences for interpretation of chemical bonding.

\end{itemize}

\subsection{Kinetic stress}
\label{sec:stress-kinetic}

In the present work, we are especially interested in the kinetic contribution to the stress field. 
Even though the tensor expressions for the kinetic stress are more complicated to write out than the scalar expressions for the energy, they are analogous expressions in terms of first and second derivatives.
The two expressions for the kinetic energy in \refeqs{eq:gen-T-1} and \ref{eq:gen-T-2} become:
 %are the same except the scalar combination of the derivative are replaced by tensors:
\begin{eqnarray}
\Psi \nabla^2 \Psi & \rightarrow & \Psi^* \nabla_{\alpha}\nabla_{\beta} \Psi \nonumber \\
 |\nabla \Psi|^2 & \rightarrow & \nabla_{\alpha} \Psi^* \nabla_{\beta} \Psi,
 \label{eq:stress-1-2}
\end{eqnarray}
and the factor $1/2$ in the kinetic energy is replaced by $-1$ in the kinetic stress.
%, which is derived below in \refsec{sec:relation}.
%The same changes apply to generalize the rest of the equations for kinetic energy to the forms for kinetic stress.
The same reasoning as done for the energy in \refsec{subsec:ke-xc} leads to the conclusion that the kinetic stress field
can be separated into a part expressed
in terms solely of the density and a part due to exchange and correlation,
\begin{equation}
\sigma^t_{\alpha \beta}(\vecr)   = \sigma^{t,n}_{\alpha
\beta}(\vecr)  + \sigma^{t,xc}_{\alpha \beta}(\vecr),
 \label{eq:stress-field-tau-xc-n}
\end{equation}
where $\sigma^{t,xc}_{\alpha \beta}(\vecr)$ is a unique
tensor field defined at every point independent of the choice
of kinetic energy operators. Similarly, there is no effect of adding a function with zero divergence so long as such terms are required to be physical, \ie they result from exchange and correlation among particles, in which case they depend only on relative coordinates and extend only over a finite region. 
%\marginpar{Is txc correlated motion across a plane?}
%\marginpar{Give Eq. for xc term?}

The issues of the choice of the kinetic energy operators for the stress field occur in the density-dependent terms,
% $s(\vecr)$
% %$s(\vecr) = \sqrt{n(\vecr)}$
% which is 
which are equivalent to a one particle equation.  This has been considered in many works \cite{RMM-nielsen85,godfrey88,filippetti00,rogers02,anderson-JCPA114-8884} and references in those papers, with the result which can be expressed in terms of $s(\vecr)$,
\begin{eqnarray}
\sigma^{t,n}_{\alpha \beta}(\vecr)  =
& - & N \left[\nabla_{\alpha} s(\vecr)\nabla_{\beta} s(\vecr)
\; -  \; s(\vecr) \nabla_{\alpha}\nabla_{\beta} s(\vecr)\right] \nonumber \\
& - & \frac{1}{4}  B \left\{ \delta_{\alpha
\beta}\nabla^2 n(\vecr) - \nabla_{\alpha}\nabla_{\beta}n(\vecr)\right\}.
 \label{eq:stress-field-tau-s-text}
\end{eqnarray}
which can also be expressed as
\begin{eqnarray}
\sigma^{t,n}_{\alpha \beta}(\vecr)  =
& - & \frac{1}{4} \left[ \frac{\nabla_{\alpha}n(\vecr) \nabla_{\beta}n(\vecr)}{n(\vecr)}
- \nabla_{\alpha}\nabla_{\beta}n(\vecr) \right] \nonumber \\
& - & \frac{1}{4}  B \left\{ \delta_{\alpha \beta}\nabla^2 n(\vecr) - \nabla_{\alpha}\nabla_{\beta}n(\vecr)
\right\}.
 \label{eq:stress-field-tau-n-text}
\end{eqnarray}
The first line in \refeq{eq:stress-field-tau-s-text} was derived long ago using momentum conservation \cite{schr27, fock30,pauli33} and more recently by others \cite{epstein-10-1063,kugler67,godfrey88,rogers02}. A lucid explanation for the form of the kinetic energy operators can be found in the paper by Godfrey\cite{godfrey88}, where equations 2 and 3 show explicitly that it is the average of the two forms of the kinetic energy operators, \ie $A=1/2$ in \refeqs{eq:t-boltzmannon} and \ref{eq:t-boltzmannon-density}. In \refsec{sec:stress-simple} we derive this form in a way that brings out the interpretation and the difference from the kinetic energy density.   %as given in \refeq{eq:stress-field-tau-s-text} and discussed in \refapp{app:stress-kinetic}.
% which is different from the energy density in \refsecs{sec:ke-density} to \ref{sec:e-density-dft}. 

The second line scaled by the parameter $B$ is the addition of a term\cite{godfrey88,rogers02} with zero divergence.  Even though one can add any function with zero divergence, Godfrey\cite{godfrey88} has argued this is the only reasonable, physically motivated choice. 
The relation \refeq{eq:stress-field-tau-n-text} in terms of the density shows why this term is a cause for possible ambiguity, since it can be combined with similar terms on the first line to create different expressions depending on the choice of the parameter $B$. Here we propose a way to deal with this issue by first considering one-dimension, where there is no ambuguity, and then the generalization to higher dimensions.  
% Examples of the effect of this term in \refapp{app:stress-curl-terms} lead to the conclusion that $B=0$ is only reasonable. With this choice, all the terms in the expressions for the stress field are uniquely determined.  

 % In this paper we provide additional ways to understand physical basis for the kinetic operators in the first line of \refeq{eq:stress-field-tau-s-text}.  They are different from the form for the kinetic energy density; this is not an ambiguity but a fundamental difference between kinetic energy and kinetic stress on a local microscopic scale. 

\subsubsection{Single particle in one dimension}
\label{sec:stress-simple}

The form for the kinetic energy operators can be determined by considering a one dimensional problem, where the stress is uniquely defined by the forces. This is shown explicitly in \refeqs{eq:stress-field-tau-s-text} and \ref{eq:stress-field-tau-n-text} since the coefficient of $B$ is zero in one dimension. 
It is straightforward to carry out the proof for a single particle in one dimension, where
the stress is a scalar $\sigma(x) = \sigma_{xx}(x)$, and problem is defined by the potential $V(x)$ and wave function $\psi(x)$. To simplify the notation we define $\psi'= d\psi/dx$ and $\psi'' = d^2\psi/dx^2$ so the \Schr\ equation can be written as
$-\frac{1}{2} \psi'' + V \psi = E \psi$. 

It is illuminating to follow the approach of Feynman (\cite{feynman39a} page 20), but here restricted to one dimension. The steps are to propose a form for the stress
\begin{equation}
    \sigma = \frac{1}{2} \left[ \psi \psi'' - (\psi')^2 \right],
    \label{stress-1d}
\end{equation}
and to show that it satisfies the divergence equation \refeq{eq:gen-stress-force}.  It is straightforward to show that 
\begin{equation}
    \frac{d \sigma}{dx} = \frac{1}{2} \left[ \psi \frac{d}{dx}\psi'' - \psi' \psi'' \right].
    \label{div-stress-1d}
\end{equation}
Substituting $\psi'' = 2(V-E) \psi$ in this equation we find that the terms proportional to $\psi'$ and $V$ cancel, leaving only one term involving the gradient of $V$ satisfying the divergence equation
\begin{equation}
    \frac{d \sigma}{dx} = - \psi^2 \frac{dV}{dx},
    \label{force-stress-1d}
\end{equation}
where the right hand side is the force acting on the electrons at point $x$ with density $n(x) = \psi(x)^2$

The conclusion is that the function in \refeq{stress-1d} is the only combination of derivatives of the wave function that leads to the correct divergence relation \refeq{force-stress-1d}. It is the kinetic stress and it is not a measure of the kinetic energy.  In fact, the proof involves the condition that $\psi$ satisfies the \Schr\ equation, where the kinetic energy is given by $-\frac{1}{2} \psi \psi''$, \ie the same as the kinetic energy density identified in {\refsec{sec:e-density-dft} and different from the stress field in \refeq{stress-1d}.

The expression for the stress is illustrated by the examples of square wells \refapp{app:stress-square-wells}.  It is immediately clear that \refeq{stress-1d} is the only form that correctly describes the stress field inside and outside the well.  The stress for the particle inside the well balances the force provided by the walls of the square well.  
% An illuminating result is that the stress is zero in the classically forbidden region.  

\subsubsection{Higher dimensions}
\label{sec:stress-2-3d}

There are two aspects in the generalization to dimensions greater then 1. One is that  the stress must involve the same combination of derivatives as found for one dimension, \ie the average of the two forms for the kinetic energy operators in \refeq{eq:stress-field-tau-s-text}; otherwise it would not reduce to the one-dimensional form we have derived.

The other is the possibility of adding a function with zero divergence, the second line in \refeq{eq:stress-field-tau-s-text} and \ref{eq:stress-field-tau-n-text} which is scaled by the parameter $B$.  
 Although we cannot rigorously derive a value of $B$, we can explore the consequences of various choices. In \refapp{app:stress-curl-terms} is the example of a two-dimensional square well.  For a long narrow well the result with any non-zero $B$ is a stress field is dominated by a term that varies rapidly in a region where there are no forces. Although this is allowed in a mathematical sense, it is problematic for physical interpretation, and  it is logical to require $B=0$ as the only acceptable choice. Since $B$ is a parameter independent of the \ham, we conclude that $B=0$ is the only reasonable choice for all systems.
 
With this choice the kinetic stress is uniquely given by the first lines of \refeq{eq:stress-field-tau-s-text} and \ref{eq:stress-field-tau-n-text} plus the exchange-correlation terms in a many-body system.  The derivation has used the form of the stress field specifically for the quantum electronic structure problem, and it is important to point out that the conclusions may not apply to other problems.

%VVVVVVVVVVVVVVVVVVVVVVVVVVV

% For the full many-body problem, the total stress field is given by the addition of the terms due to exchange and correlation
% in \refeq{eq:stress-field-tau-xc-n-text} and contributions due to interactions.   As discussed in \refapp{app:stress-derivations}, these terms are well-defined independent of the choice of kinetic energy operators.
% The interactions terms also are well-defined in terms electric fields in the Maxwell formulation, but are different from the corresponding terms in the energy density.

\subsection{Stress due to interactions}
\label{sec:stress-interactions}

Since stress is related to external forces on the particles and internal forces between particles, the natural formulation for a stress field is in terms of the interaction field instead of potentials acting on particles. (See \cite{RMM-nielsen85,rogers02} and earlier references given there.) 
The interactions contribution to the stress $\sigma^{int}_{\alpha \beta}(\vecr)$
can be divided into terms due to the external  and mean-field terms plus others that involve the exchange and correlation among the particles; however, each is different from the expressions give in \refsecs{sec:pe-density} and \ref{sec:e-density-dft} for the energy.    
% Just as for the difference in the kinetic terms, this is not an inconsistency but rather an essential difference in energy and stress.

For Coulomb interactions, 
% the energy density at point $\vecr$ is given in terms of electric fields at $\vecr$ \cite{lan60,jac62}. 
% %Dec - commented out \marginpar{Mention non-Coulomb cases?  Kugler NM}
% %They considered a local density approximation, in which case the total potential stress is the given by
The mean-field terms are given by \cite{lan60,jac62}
\begin{equation}
\sigma^{Maxwell}_{\alpha \beta}(\vecr)   = \frac{1}{4 \pi} \left[
\vecE_{\alpha}(\vecr) \vecE_{\beta}(\vecr) - \frac{1}{2} \delta_{\alpha
\beta}\vecE_{\gamma}(\vecr)
\vecE_{\gamma}(\vecr)\right].
 \label{eq:stress-Maxwell}
\end{equation}
Regarding the contributions of correlation, the expressions in terms of electric fields 
%full \mb\ form 
is given by Eq. B1 of NM \cite{RMM-nielsen85} as the
expectation value of the electric field operators
\begin{equation}
 \sigma^{int}_{\alpha \beta}(\vecr)  =
\frac{1}{4 \pi} \langle \Psi \vert \hat{\vecE}_{\alpha}(\vecr)
\hat{\vecE}_{\beta}(\vecr)  - \frac{1}{2} \delta_{\alpha
\beta}\hat{\vecE}_{\gamma}(\vecr)\hat{\vecE}_{\gamma}(\vecr) \vert
\Psi \rangle
 \label{eq:maxwell-stress-mb}
\end{equation}
which includes fluctuations in the electric fields in addition to
the average electric fields. The mean square fluctuations are the natural formulation of a stress field since stress is due to the equal and opposite forces between particles across
a unit plane due to fields that cross the plane, as described by NM \cite{RMM-nielsen85}.  

% Note that the expressions involve only the fluctuations of the fields at each point $\vecr$, not correlated fluctuations at different points. 

% This is quite different from 
% the considerations for the energy in
% \refsecs{sec:mb-system-general} and \ref{sec:e-density-dft}, where it was argued that the minimization
% conditions in terms of the density led to expressions in terms of potentials acting on particles
% with density $n(\vecr)$.
% %Dec 5 - commented out - \marginpar{Modified xc part August 8, 2024}

% Nevertheless there is a close relation to the correlation energy defined in \refsecs{sec:mb-system-general} and \ref{sec:e-density-dft}.  In a nonrelativistic problem, the fluctuations in the electric fields are actually caused by the fluctuations in the positions of the particles. The relations are given in  appendix C of NM \cite{RMM-nielsen85}, where it is clear that the result is due to correlation of particles, \ie\ the exchange-correlation hole, but the energy is associated spatially with the field instead of the potential at the particle positions. 

The stress field can be very different from the expression for correlations in terms of potentials at the positions of the particles; for example, due to the long-range interaction the stress field is non-zero in regions where there are no particles.  Several examples are described in appendix C of NM, such as the quantum fluctuations of the electric field of a polarizable object, which decay as $1/r^6$; these are the fields that give rise to the London dispersion forces and the long range part of van der Waals interactions.

% For two polarizable units the results can be described in terms of the non-local correlation of electrons on the two units; the resulting energies are the same, but the distribution in space of the energy is different and the interpretation is different, where the formulation in terms of fields shows clearly the actual basis for the interaction. 
% %

\subsection{Stress in the Kohn-Sham approach} 
\label{sec:stress-KS}

The first point to emphasize is that the Kohn-Sham calculation is not changed; stress is a property of the ground state which is calculated after the Kohn-Sham calculation is converged.  For the calculation of stress the kinetic term involves the average of the two forms, even though that is different from what was done in the Kohn-Sham calculation. Nevertheless, this does not present great difficulties in practice; as noted following \refeq{eq:t-boltzmannon-density}, any form can be used so long as one adds a term that involves the laplacian of the density.

 The contributions from interactions are included in terms of electric fields.  The mean field terms  $\sigma^{Maxwell}_{\alpha \beta}(\vecr)$ in \refeq{eq:stress-Maxwell} can be found by using the relations  $\vecE(\vecr) = \nabla V_{tot}(\vecr)$ and $- \nabla^2 V_{tot}(\vecr) = n_{tot}(\vecr)$, where $V_{tot}$ and $n_{tot}$ are the total potential and charge density including both electrons and nuclei.  However, the exchange correlation term is problematic because it involves the calculation of the electric fields due to the correlations.   In principle, it can be calculated by a many-body calculation, but it is not given directly by the Kohn-Sham correlation function.  To the knowledge of the authors, this has not been addressed in previous work. 
 
 In the local approximation the exchange-correlation functional is taken to be a local function of the density, in which case the spatial dependence would be the same no matter whether the form in terms of potentials or electric fields is used. In this case the stress is the form given by given by Godfrey\cite{godfrey88} and Rogers and Rappe\cite{rogers02}  %
 %Jan 8 2025 corrected factor from 1/4 to 1/2
\begin{eqnarray}
\sigma_{\alpha \beta}(\vecr) &=& \frac{1}{2} \sum_i\left[ \psi_i(\vecr)\nabla_{\alpha} \nabla_{\beta} \psi_i(\vecr) - \nabla_{\alpha}\psi_i(\vecr)\nabla_{\beta}\psi_i(\vecr) \right] \nonumber \\
&+& \sigma^{Maxwell}_{\alpha \beta}(\vecr) + \delta_{\alpha \beta} \sigma^{LDA}(\vecr),
\label{eq:stress-KS}
\end{eqnarray}
 where the $\psi_i(\vecr)$ are Kohn-Sham orbitals.  The last term is a local pressure since the exchange-correlation stress is diagonal in the LDA  and is given by the volume dependence of the LDA functional evaluated at point $\vecr$. 
An explicit expression for the local pressure has been proposed by Tsirelson, \etal\ using the LDA functional and an approximation for the kinetic energy in terms of the density\cite{tsirelson:px5014}.

\section{Relation of local pressure and energy density fields}
\label{sec:relation}

In the previous sections one of the main results is that the expressions are different for the energy density and the stress fields.   
Nevertheless, in much of the literature there is a different choice for the kinetic energy density that is derived from the local pressure 
\begin{equation}
t^{PS}(\vecr) = -\frac{1}{2} Tr[ \sigma^t_{\alpha \beta}(\vecr)] = \frac{3}{2} p(\vecr).
 \label{eq:energy-pressure}
\end{equation}
often ascribed to Pauli \cite{pauli33}, \Schr\ \cite{schr27} and others. According to \cite{anderson-JCPA114-8884} the ``ambiguity in the local kinetic energy can be considered
to be inherited from the ambiguity in the local electronic stress tensor'' given in \refeq{eq:energy-pressure}.

The resolution is that stress and energy are physically different quantities and we must analyze carefully the relation between them. For a homogeneous system the value of $t^{PS}$  in \refeq{eq:energy-pressure} is equal to the kinetic energy density $t$ in \refeq{eq:t-B-F}.  In this case $t_n = 0$, since the density is constant, and both kinetic energy and pressure are due solely to exchange and correlation,   which are independent of the choice of the form of the kinetic energy operators.  This is a general relation for a homogeneous many-body system and it not restricted to independent-particles. 
% For example, for a homogeneous gas of non-interacting fermions, the energy and pressure are due strictly to the kinetic energy for fermions due to the exclusion principle.  

However, this does not apply as a function of position in an inhomogeneous system. As we have shown, the kinetic energy density $t_n(\vecr)$ in density functional theory involves $-\frac{1}{2}s(\vecr)\nabla^2 s(\vecr)$, whereas the local pressure $t^{PS}(\vecr)$ in \refeq{eq:energy-pressure}  involves $\frac{1}{4}[s(\vecr)\nabla^2 s(\vecr) - |\nabla s(\vecr)|^2]$.  
% The assumption that the quantity $t^{PS}(\vecr)$ defined in \refeq{eq:energy-pressure} is a local energy density in an inhomogeneous system can be viewed as an approximation analogous to the Thomas-Fermi local approximation.  In this sense each is a local approximation that the energy density for the homogeneous system applies locally in an inhomogeneous system.  
The correct expression requires the analysis in terms of the ground state wave functions to derive $t(\vecr) = t_n(\vecr) + t_{xc}(\vecr)$, $\sigma^t_{\alpha \beta}(\vecr)   = \sigma^{t,n}_{\alpha
\beta}(\vecr)  + \sigma^{t,xc}_{\alpha \beta}(\vecr)$, and the corresponding interaction contributions.  

% In one dimension, the analysis in \refsec{sec:stress-simple} and \refapp{app:stress-1d-square-well} shows clearly the difference between kinetic stress and energy.  The only ambiguity is in higher dimensions where there is a possible addition of a field with zero divergence; however, we have shown that any such addition can lead to unreasonable results.   

The conclusion is that one should simply recognize the difference between energy and stress.  For example, in the analysis of chemical bonding, it is in some ways irrelevant if one chooses to use the local pressure $p(\vecr) = -\frac{1}{3} Tr[ \sigma^t_{\alpha \beta}(\vecr)] $ or the quantity $t^{PS}(\vecr) =\frac{3}{2} p(\vecr)$ defined by \refeq{eq:energy-pressure}, since they are proportional to one another.  However, energy and stress have different physical meanings and
%, in the opinion of the author, is that it is 
stress (or pressure) is the appropriate quantity for analyzing bonding.  This is actually what is done in many papers, even if it is called energy density.
The undergraduate thesis of Feynman\cite{feynman39a} gives the physical picture clearly.  Using the example of the water molecule and ``angular stiffness'', he gave a picture of stresses across a bisecting plane and says ``A real answer to such problems could, of course, only come with considerations of the stresses in space in various regions near the molecule.''

\section{Conclusions}
\label{sec:Conclusions}

This paper has considered energy density and stress
fields in a nonrelativistic system of many interacting particles.
%, in particular electrons in the presence of external potential such as nuclei.
This is a subject with a long history of controversy and
confusion concerning whether or not such fields can be defined
uniquely, and how they can be used, as summarized, \eg in \cite{anderson-JCPA114-8884}. 
The kinetic energy densities are different depending on the choice of the forms $\frac{1}{2}|\nabla \Psi|^2$, $-\frac{1}{2} \Psi \nabla^2 \Psi$ or a linear combination of the two forms. There are also two ways to describe the interactions as potentials acting on the particles or fields between the particles.  For Coulomb interactions the choices are an energy density in terms of potentials or in terms of electric fields.  
% For both kinetic and interaction energies there is the issue of how to include effects of exchange and correlation.

A principal result of the present work is the division of the problem into a part that depends only on the particle density $n(\vecr)$ plus a part that includes all affects of exchange and correlation. For the kinetic energy terms, it is shown in \refsec{sec:ke-density} 
%and \ref{sec:stress-field} 
that all contributions 
%to the kinetic energy density and stress fields 
due to exchange and correlation among the particles are unique and well-defined at each point $\vecr$, independent of the choice of form of the kinetic energy expressions.
All issues of nonuniqueness can be cast in terms of the density and its derivatives, which correspond to a single particle problem with wave function $s(\vecr) = \sqrt{n(\vecr)/N}$. 
% A definite form for these terms can be determined only if there are added constraints or other reasons, and the analysis is greatly facilitated since the issues are the same as in single-particle problems.  
% %The are analogous issues for the interaction energies, with some additional considerations. 
For the interaction terms, there are different representations in terms of fields or potentials and one must identify the proper choice.  In each case, there is an  division into mean-field and exchange-correlation terms, with issues of nonuniqueness confined to the mean-field terms. .   

Construction of a well-defined energy density $e(\vecr)$ is based on the fact that energy plays a special role since it is the quantity that determines the ground state wave function through the variational principle. 
%In this capacity the appropriate choice of an energy density $e(\vecr)$ is one that applies for densities in a neighborhood of the ground state so that it facilitates the minimization of the energy. 
This is the topic of \refsec{sec:e-density-dft}.  For the kinetic energy terms this leads directly to the kinetic energy operator in the form
$-\frac{1}{2} s \nabla^2 s$ and interactions considered as potentials acting on the particles, as in the \Schr\ equation. However, this is not merely a restatement of the minimization principle for the original many-body problem; it is the construction of an energy density $e(\vecr)$ that is equilibrated in the ground state. As discussed in 
 \refsec{subsec:dft} this leads directly to density functional theory in a way similar to the Levy-Lieb approach. This formulation suggests ways to divide the problem into with different methods to equilibrate the density and to find the exchange-correlation density. The reasoning carries over directly to the Kohn-Sham approach.

In contrast, a stress field $\sigma_{\alpha \beta}(\vecr)$  is related to forces and satisfies the divergence equation \refeq{eq:gen-stress-force}, which has the well-known problem that one can always add a field with zero divergence. In \refsec{sec:stress-field}  we propose that for a quantum system in the ground state the form is rigorously determined by considering a one dimensional system and generalizing to any dimension. 
The result is that each contribution is well-defined and each is different from the corresponding term in the energy density. The kinetic stress involves the combination $\frac{1}{2}[s\nabla^2 s - |\nabla s|^2]$, as was derived in many previous works starting in the early days of quantum mechanics by
\Schr\ \cite{schr27}, Fock\cite{fock30} and Pauli\cite{pauli33} and later by others such as in \cite{epstein-10-1063,kugler67,godfrey88,rogers02}. 
In this paper, the form is derived using a simple analysis in \refsec{sec:stress-simple} and pedagogical models \refapp{app:stress-square-wells}. The interaction terms are properly described by forces between particles instead of potentials acting on particles. For the Coulomb interaction this leads to the Maxwell formulation for the energy-stress tensor in terms of electric fields.

Energy density and stress fields have been used to interpret chemical bonding despite lack of agreement on the meaning and the actual forms, as discussed in the Introduction. In particular, in \refsec{sec:relation} is an energy density derived directly from the local pressure,  $t^{PS}(\vecr) = -\frac{1}{2} Tr[ \sigma^t_{\alpha \beta}(\vecr)] = \frac{3}{2} p(\vecr) $ given in \refeq{eq:energy-pressure}, a well-known result derived in Refs. \cite{schr27,fock30,pauli33,epstein-10-1063,kugler67,anderson-JCPA114-8884}. The conclusion in \refsec{sec:relation} is that this is indeed the correct result in a homogeneous system. But for the problem of spatially varying fields, the energy and stress are fundamentally different and $t^{PS}(\vecr)$ on the left hand side of \refeq{eq:energy-pressure} is not the same as a kinetic energy. It can be viewed as a local approximation analogous to the Thomas-Fermi approximation. The important result of the present work is that local pressure $p(\vecr)$ is a well-defined quantity, and we argue that it is the appropriate quantity to use for interpretation of chemical bonding.
%, and it is just a matter of notation whether it is called $p(\vecr)$ or $t^{PS}(\vecr)$.   

Finally, it should be emphasized that many integrated quantities are not subject to any of the issues of nonuniqueness discussed in this paper.
These include the total energy and forces in a finite system, the energy per cell in a crystal,
the force on any atom,
the macroscopic stress in an extended system, the surface energy or stress
in a solid,
%\cite{che92,che92a,rapcewicz98,filippetti00}.
and the kinetic energy integrated over the region around an atom defined
by the condition $\nabla n(\vecr) = 0$ which is used to define ``atoms in molecules''\cite{bader90,bader91}.   Since the results are invariant to
the choice of the form of the energy or stress field, one is free to use whatever
expression is most advantageous for calculation of any of these quantities.

\section{Acknowledgements}
\label{sec:Acknowledgements}

%\marginpar{July25 - Added Pavanello}

The authors are indebted to M. H. Cohen, O. H. Nielsen, R. J. Needs and D. H. Vanderbilt  for continuing interactions on this topic over the years;
K. Burke, A. Cancio, D. M. Ceperley, J. Krogel, P. Ordejon, A. M. Rappe, L. Shulenberger, and E. Stechel for stimulating discussions; M. Yu for our work together on energy density; and M. Pavanello for discussions about ``orbital-free'' methods and pointing out the relation to the exact factorization approach. This work was
948 supported by the National Science Foundation under Grant
949 No. DMR-1006077, and through the Materials Computation
950 Center at UIUC (NSF Grant No. DMR-0325939), and by the U.S.
     Department of Energy, Division of Materials Sciences under
     Award No. DEFG02-91ER45439, through the Frederick Seitz
     Materials Research Laboratory at the University of Illinois at
     Urbana-Champaign.  Parts of this work were done at the Aspen
     Center for Physics and at Lawrence Livermore National Laboratory.

\appendix

\section{Examples of square wells}
\label{app:stress-square-wells}

\subsection{Square wells in one dimension}
\label{app:stress-1d-square-well}

In order to illustrate that the form of stress in \refeq{stress-1d} is the only correct expression, we can define a general combination of derivatives
\begin{equation}
    \sigma^A = \left[ A\psi \psi'' - (1-A)(\psi')^2 \right],
    \label{stress-1d-test}
\end{equation}
and show that $A$ must be $\frac{1}{2}$. 
A convenient example is a one dimensional box with of length $L$, constant potential $V=0$ inside the box, and an infinite barrier outside.  In any region where the potential is constant,
the stress must be constant as required by \refeq{eq:gen-stress-force}.   The ground state wave function is $\psi(x) = \sqrt{2/L} \, cos(\pi x/L)$ where the origin is chosen at the center.
%  All potential terms vanish and the total stress field is
% $\sigma^{t,n}_{xx}(x)$ given in \refeq{eq:stress-field-tau-s-text} or \refeq{eq:stress-field-tau-s}.
% %or \ref{eq:stress-field-tau-n}.   
It is a simple matter to carry out the derivatives 
%in \refeq{eq:stress-field-tau-s} 
to find the stress,
\begin{equation}
\sigma_{xx}(\vecr)  = - N \frac{2\pi^2}{L^3} \left[
A sin^2(\pi x/L) + (1-A) cos^2(\pi x/L)\right].
 \label{eq:stress-field-tau-s-1D-box}
\end{equation}
The only value of $A$ for which the stress is constant is $A=1/2$, \ie\ the ``Pauli-Schr\"{o}dinger-Epstein'' form which is the average of the two characteristic choices of the kinetic energy operators.

\begin{figure}
\begin{center}
\includegraphics[width=8.0cm]{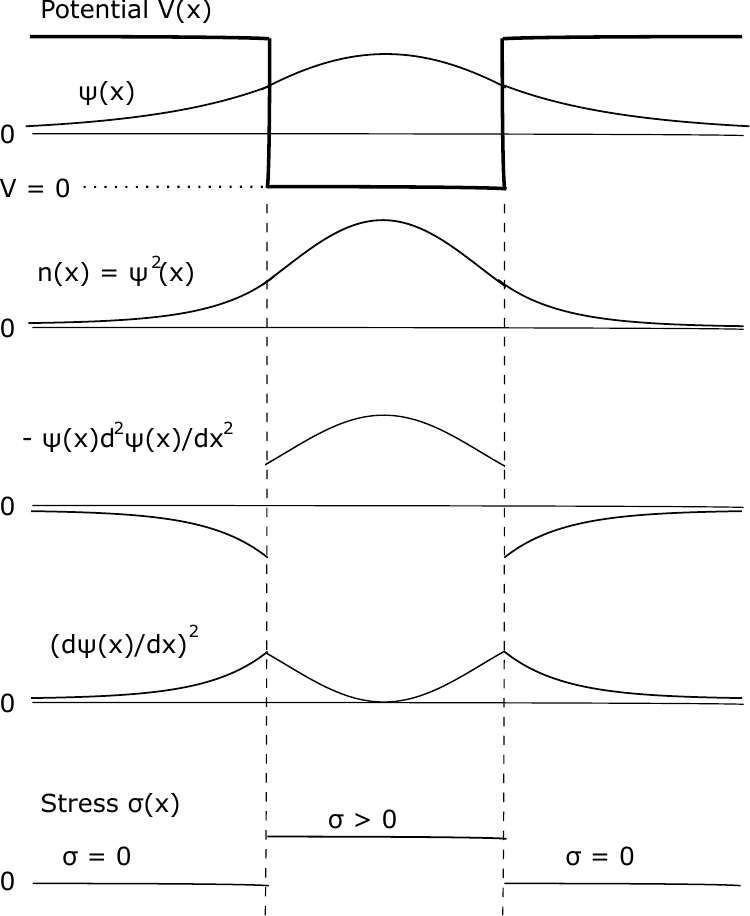}
\caption{
Example of a finite square well, shown in the top line along with a characteristic wave function $\psi(x)$ that is sinusoidal inside the well and decays exponentially outside the well. The third and fourth lines show the two forms of the kinetic energy density.  The bottom line shows the stress for the case $A=1/2$ which is the average of the two forms for the kinetic energy density.  Any other combination leads to an unphysical result for the stress field.   
}
\label{fig-square-well}
\end{center}
%\end{framed}
\end{figure}

The choice of kinetic operators is even more illuminating if we consider a well
with a potential that is finite and a wave function that decays exponentially outside the well,
$\psi(x) \propto e^{-ax}, \; x>L/2$,
as shown in Fig. \ref{fig-square-well}. The figure shows the two forms of the kinetic energy, one positive outside the well and the other negative, but with the same magnitude.  
The correct result is that the stress must be zero outside the well because  it is zero for large $x$ where $\psi(x) \rightarrow 0$ and it must have the same value everywhere outside the well since the potential is constant.  The stress is given by \refeq{stress-1d-test},
\begin{equation}
\sigma(x) \propto
-  a^2 e^{-2ax} \left[ A - (1-A) \right],
 \label{eq:stress-psi-zero}
\end{equation}
which is zero only if $A= 1/2$.  
%The fact that the stress is zero in the classically forbidden region has interesting consequences discussed in \refsec{sec:relation}.

\subsection{Arguments for choice of the arbitrary divergenceless terms}
\label{app:stress-curl-terms}

In order to examine the effect of the parameter $B$, we need to consider a problem in dimension greater than 1.
A convenient choice is a two-dimensional rectangular box with
dimensions $L_x,L_y$. The ground state wave function is $s(\vecr)
= \sqrt{2/L_x} \, (cos(\pi x/L_x) \sqrt{2/L_y} \,cos(\pi y/L_y)$.
%and $n(\vecr) = \frac{2}{L_x} cos(\pi x/L_x)^2 \frac{2}{L_y}cos(\pi y/L_y)^2$,
%where the origin is chosen at the center.
If the box is finite but very long in the $\hat{y}$ direction, $L_y >> L_x$, the derivatives in the $\hat{y}$ direction are much smaller than in the $\hat{x}$ direction.  The stress $\sigma_{xx}(x,y)$ is
scaled by the density as a function of $y$ but it still is constant as a function of $x$ and has nearly the same form as in one dimension. The stress in the $y$ direction is given by
%is greatly affected by the has a large contribution from $\sigma_{yy}(x,y)$,
\begin{eqnarray}
\sigma_{yy}(x,y) &=& - N \frac{4}{L_xL_y} \left[ \frac{\pi^2}{L_y^2} cos^2(\pi x/L_x) \right. \\
 & & \left. + \frac{B}{2}\frac{\pi^2}{L_x^2}(sin^2(\pi x/L_x) - cos^2(\pi x/L_x)) \right], \nonumber
 \label{eq:stress-yy}
\end{eqnarray}
where we have used $A=1/2$.  If $B \neq 0$, the last term is large and oscillates as a function of $x$.
In principle such a term is allowed, but it forces an interpretation of the quantum behavior based on an arbitrary parameter. In practice, it is problematic since it varies rapidly as a function of $x$ even though there are no forces anywhere in the well, and it is large at the boundaries $\pm L_x/2$ even though there are no forces in the $\hat{y}$ direction due to the boundaries.  Thus $B \equiv 0$ is the only reasonable formulation for the stress field.  This has been termed the ``minimal'' choice and $B = 0$ is used in many analyses, as discussed in \cite{tsirelson:px5014}.

Another way to reach this conclusion is that $B \equiv 0$
is the only choice that respects the separability of this problem.
The \Schr\ equation has independent variations in each direction, and $B \equiv0$ is the only
choice for which the stress is separable.  Furthermore, this implies there should
never be any ambiguities because the problem breaks into a product
of one-dimensional problems and there is no ambiguity in the kinetic
stress in one dimension.

\bibliography{ELSGP,RMM-68-86-articles,BIBFILE-COMBINED-2017,NEW-REFS-2017,density-paper-new-refs-2024}

%\end{multicols}
\end{document}